\documentclass[12pt]{article}
\usepackage{physics}
\usepackage{epsf}
\usepackage{aas_macros}
\usepackage{amssymb}
\usepackage{comment}
\usepackage{amsmath}
\usepackage{braket}
\usepackage{mathrsfs}
\usepackage{mathtools}
\usepackage{array}
\usepackage{fancyhdr}
\usepackage[dvipdfmx]{graphicx}
\usepackage{color}
\usepackage{cite}
\usepackage{bm}
\usepackage{url}
\usepackage{fancyhdr}
\usepackage{mathtools}
\usepackage{calc}
\usepackage[colorlinks,citecolor=blue]{hyperref}
\usepackage[hang,small,bf]{caption}
\usepackage[subrefformat=parens]{subcaption}
\usepackage{scalerel}
\usepackage[usestackEOL]{stackengine}

\renewcommand{\thefootnote}{\#\arabic{footnote}}

\renewcommand{\thefootnote}{\fnsymbol{footnote}}
\def\thefootnote{\fnsymbol{footnote}}
\newcommand{\FP}{\mathop{\mathrm{FP}}}

\makeatletter

\@addtoreset{equation}{section}
\makeatother

\def\be{\begin{equation}}
\def\ee{\end{equation}}
\def\ben{\begin{eqnarray}}
\def\een{\end{eqnarray}}

\begin{document}

\begin{center}

\vskip .75in

{\Large \bf A universal connection between lens density profiles and low-frequency wave optics in gravitational-wave lensing}

\vskip .75in

{\large
So Tanaka$\,^1$,  Teruaki Suyama$\,^1$
}

\vskip 0.25in

{\em
$^{1}$Department of Physics, Institute of Science Tokyo, 2-12-1 Ookayama, Meguro-ku,
Tokyo 152-8551, Japan
}

\end{center}
\vskip .5in

\begin{abstract}

We investigate the low-frequency behavior of the amplification factor in gravitational lensing and explore how it encodes information about the density profile of the lensing object. We derive the low-frequency expansion of the amplification factor under the Born approximation for a broad class of projected density profiles. For spherically symmetric profiles that decay faster than any power law at large distances, we derive a systematic expansion of the amplification factor in powers of frequency, with the logarithmic dependence appearing only in the leading term, and show that each expansion coefficient is determined by a finite set of moments of the density profile. We then extend the analysis to profiles with power-law tails and demonstrate that such profiles induce additional non-analytic frequency dependences, including fractional powers and logarithmic terms, which directly reflect the asymptotic behavior of the density distribution. Furthermore, we investigate the effects of non-sphericity and show that contributions from the quadrupole moment appear only as higher-order corrections relative to the spherically symmetric component in the low-frequency regime. Finally, we investigate the validity of the Born approximation in the low-frequency expansion. We derive a criterion for the maximum order of the low-frequency expansion up to which the Born approximation remains dominant over the post-Born corrections.

\end{abstract}

\renewcommand{\thepage}{\arabic{page}}
\setcounter{page}{1}
\renewcommand{\thefootnote}{\#\arabic{footnote}}
\setcounter{footnote}{0}

%===============================================

\section{Introduction}
\label{Sec: Introduction}

Gravitational lensing is a powerful tool for probing the mass density distribution in the Universe\cite{Tambalo:2022wlm,Oguri:2020ldf,Savastano:2023spl}. Gravitational lensing of light, namely gravitational lensing in the geometric-optics regime, has been extensively studied\cite{schneider2012gravitational,Bartelmann:2010fz}. In the geometric-optics regime, lensing effects are characterized by several observables, such as image positions, magnifications, and time delays, which contain information about the lensing object. Furthermore, the gravitational lensing of gravitational waves has attracted considerable attention in recent years\cite{Choi:2021bkx,Dai:2018enj, Tanaka:2023mvy, Tanaka:2025ntr,Suyama:2025gbh,Mizuno:2023pru, Nakazono:2026vei, Braga:2024pik, CarrilloGonzalez:2025gqm, Chakraborty:2024mbr, Chakraborty:2025maj, Chakraborty:2025pxt}. For long-wavelength waves such as gravitational waves, wave-optical effects, including diffraction, become important\cite{Nakamura:1997sw,Nakamura:1999uwi,Takahashi:2003ix}. The frequency dependence of wave-optical lensing effects contains information about density distributions that cannot be accessed in geometrical optics, and is therefore expected to provide a new probe of dark matter distributions\cite{Takahashi:2005ug,Tanaka:2025nfw,Kim:2025njb}.

To understand what information about the density distribution is encoded in wave-optical gravitational lensing, it is useful to study the low-frequency expansion of the amplification factor. However, such expansions have so far been derived only for a few specific lens models\cite{Tambalo:2022plm, Choi:2021bkx}, and no unified expansion formula applicable to general density profiles is known. Consequently, a general understanding of how the coefficients of the low-frequency expansion are related to the density profile has not yet been obtained.

In this work, we investigate the behavior of the amplification factor in the low-frequency regime and clarify what information it contains about the density profile. In Sec. \ref{Sec: Low-frequency expansion of amplification factor}, under the Born approximation, we derive the low-frequency expansion of the amplification factor for a broad class of density profiles. As a mathematical preparation, in Sec. \ref{Sec: Finite part prescription}, we introduce the finite-part prescription used in the calculations throughout this paper. This prescription provides a method for extracting the finite part of integrals that contain divergences. In Sec. \ref{Sec: Rapidly decaying spherically symmetric profile}, we consider the case in which the profile is spherically symmetric and decays faster than a power law at large distances, and show that the coefficients of the low-frequency expansion contain information about the moments of the profile. In Sec. \ref{Sec: Profiles with power-law tails}, we extend this result to profiles that decay as a power law and clarify the relationship between the power-law index and the terms appearing in the low-frequency expansion. Furthermore, in Sec. \ref{Sec: Effects of nonsphericity}, we investigate the contribution of non-spherical components of the profile and show that their contributions are sub-leading compared with those of the spherically symmetric components. Finally, in Sec. \ref{Sec: Applicability to realistic lens systems}, we examine the applicability of the low-frequency expansion to realistic lensing systems. Specifically, we examine up to which order the low-frequency expansion within the Born approximation can be reliably used, and evaluate this maximum applicable order for a realistic lensing system in the low-frequency regime. Throughout this paper, we use the unit $c=1\,$.

%===============================================

\section{The amplification factor in the Born approximation}
\label{Sec: Gravitational lensing system}

\begin{figure}[t]
    \centering
    \includegraphics[scale=0.5]{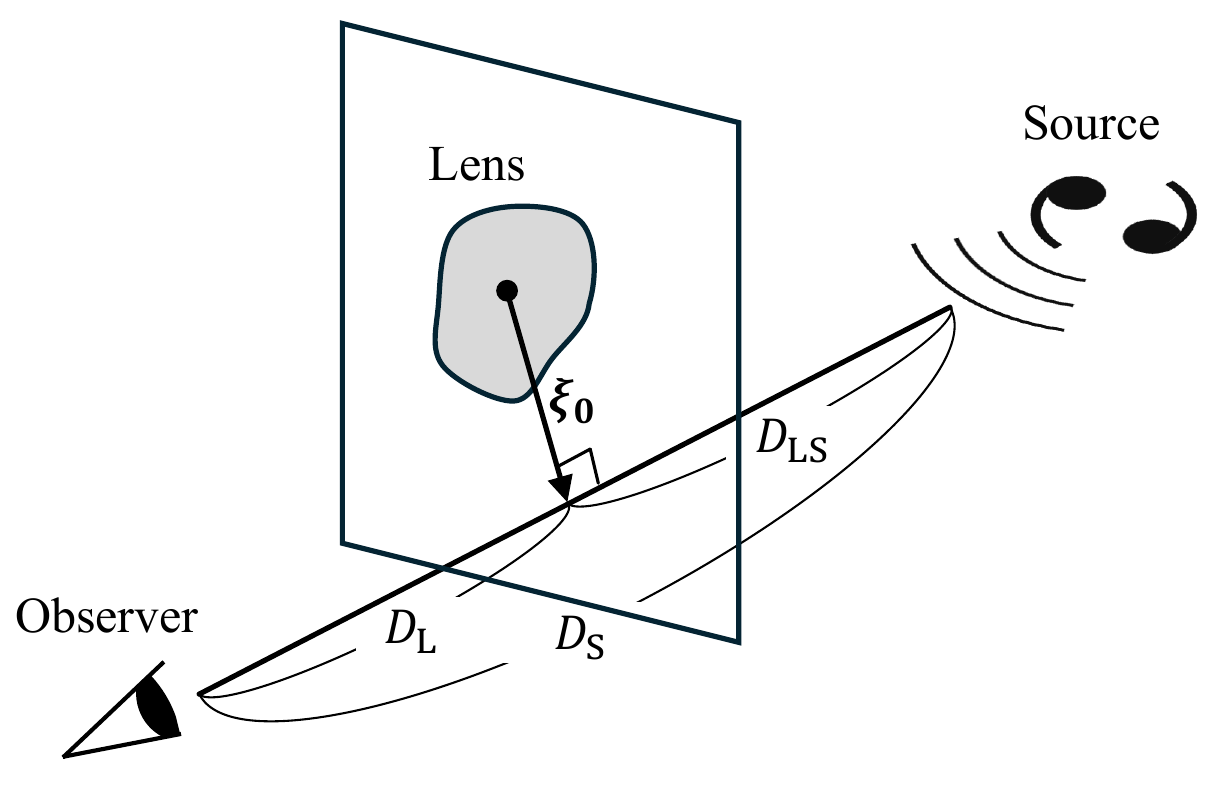}
    \caption{Schematic picture of the gravitational lensing system. $\bm{\xi}_0$ is the position vector of the intersection point between the line of sight and the lens plane, where the origin is placed on the center of mass of the lensing object. $D_{\rm L}\,$, $D_{\rm LS}$ and $D_{\rm S}$ are the distances between the observer and the lens, the lens and the source, and the observer and the source, respectively.}
    \label{Fig: gl_system}
\end{figure}
In this section, we review the derivation of the amplification factor. We consider the gravitational lensing by a single lensing object, using the thin-lens approximation\cite{Suyama:2005mx}. We also assume that only the mass density distribution in the vicinity of the line of sight contributes to the lensing effect\footnote{This assumption corresponds to $|\bm{\xi}-\bm{\xi}_0|\ll D_{\rm L}\,,\,D_{\rm LS}\,$.}. Then the amplification factor is given by\cite{Nakamura:1999uwi}
\begin{eqnarray}
    F(\omega,\bm{\xi}_0)=\frac{\omega}{2\pi id_{\rm eff}}\int d^2\xi\,\exp\qty[-i\omega\hat{\psi}(\bm{\xi})+i\omega\frac{(\bm{\xi}-\bm{\xi}_0)^2}{2d_{\rm eff}}]\,.\label{Eq: amplification factor definition}
\end{eqnarray}
Here, $\omega$ is the angular frequency of the wave undergoing gravitational lensing. $\bm{\xi}$ is the position vector on the lens plane, and $\bm{\xi}_0$ is the position of the intersection point between the line of sight and the lens plane, where the origin is placed on the center of mass of the lensing object. $\hat{\psi}(\bm{\xi})$ is the lens potential which satisfies 
\begin{eqnarray}
    \nabla_{\xi}^2\hat{\psi}(\bm{\xi})=8\pi G\Sigma(\bm{\xi})\,,
\end{eqnarray}
where $\Sigma(\bm{\xi})$ is the projected surface mass density. $d_{\rm eff}$ is the effective distance defined by
\begin{eqnarray}
    d_{\rm eff}=\frac{D_{\rm L}D_{\rm LS}}{D_{\rm S}}\,,
\end{eqnarray}
where $D_{\rm L}\,$, $D_{\rm LS}$ and $D_{\rm S}$ are the distances between the observer and the lens, the lens and the source, and the observer and the source, respectively\footnote{For simplicity, we neglect cosmological expansion throughout this paper. Nevertheless, the results obtained here can be directly applied to an expanding universe by interpreting $D_{\rm L}\,$, $D_{\rm LS}$ and $D_{\rm S}$ as the corresponding angular-diameter distance and replacing $\omega$ with $(1+z_{\rm L})\omega\,$, where $z_{\rm L}$ denotes the redshift at the lens plane.}. In addition, for later convenience, we define the Fresnel scale, which is the characteristic length scale in a gravitational lensing system, as
\begin{eqnarray}
    r_{\rm F}=\sqrt{\frac{d_{\rm eff}}{\omega}}\simeq1.3\times10^2\,\mathrm{pc}\qty(\frac{d_{\rm eff}}{10\,\mathrm{Gpc}})^{1/2}\qty(\frac{f}{\mathrm{mHz}})^{-1/2}\,,\label{Eq: Fresnel scale}
\end{eqnarray}
where $f$ is the frequency of the wave.

Let $R$ be the length scale at which the density profile decays, and define the dimensionless density profile by
\begin{eqnarray}
    \kappa(\bm{x})=4\pi Gd_{\rm eff}\Sigma(R\bm{x})\simeq\,1\qty(\frac{d_{\rm eff}}{10\,\mathrm{Gpc}})\qty(\frac{\Sigma}{1.7\times10^8M_{\odot}/\mathrm{kpc}^2})\,,\label{Eq: profile dimless}
\end{eqnarray}
where $\bm{x}=\bm{\xi}/R$ is a dimensionless position vector, and $\kappa(\bm{x})$ is assumed to decay for $x\gtrsim1\,$. 
We also define other dimensionless quantities by
\begin{eqnarray}
    \bm{y}=\frac{\bm{\xi}_0}{R}\,,\quad w=\frac{R^2}{d_{\rm eff}}\omega=\frac{R^2}{r_{\rm F}^2}\,,\quad \psi(\bm{x})=\frac{d_{\rm eff}}{R^2}\hat{\psi}(R\bm{x})\,.\label{Eq: dimless quantity}
\end{eqnarray}
The dimensionless lens potential $\psi(\bm{x})$ satisfies
\begin{eqnarray}
    \nabla_x^2\psi(\bm{x})=2\kappa(\bm{x})\,,\label{Eq: poisson equation dimless}
\end{eqnarray}
and its formal solution is given by
\begin{eqnarray}
    \psi(\bm{x})=-2\int\frac{d^2q}{(2\pi)^2}\frac{\tilde{\kappa}(\bm{q})}{q^2}e^{i\bm{q}\cdot\bm{x}}\,,\label{Eq: lens potential integral}
\end{eqnarray}
where $\tilde{\kappa}(\bm{q})$ is the Fourier transform of $\kappa(\bm{x})\,$.
From these, Eq. (\ref{Eq: amplification factor definition}) becomes
\begin{eqnarray}
    F(w,\bm{y})=\frac{w}{2\pi i}\int d^2x\,\exp\qty[-iw\psi(\bm{x})+\frac{i}{2}w(\bm{x}-\bm{y})^2]\,.\label{Eq: amplification factor dimless}
\end{eqnarray}

We now consider the geometric-optics limit $w\to\infty\,$. Since this paper focuses on the weak-lensing regime, where the lensing potential is small, the wave propagates along the almost unperturbed straight path from the source to the observer in the $w\to\infty$ limit. The corresponding Shapiro time delay is given by $t_{\rm d}=-\psi(\bm{y})\,$. Since this overall time delay is not directly observable, we remove the phase associated with $t_d$ and redefine the amplification factor according to
\begin{eqnarray}
    F(w,\bm{y})\to F(w,\bm{y})e^{iw\psi(\bm{y})}\,.\label{Eq: amplification factor redefine}
\end{eqnarray}
We then introduce the reduced amplification factor by
\begin{eqnarray}
    \eta(w,\bm{y})=F(w,\bm{y})e^{iw\psi(\bm{y})}-1\,.
\end{eqnarray}

In this paper, we focus on the low-frequency behavior of the amplification factor in the weak-lensing regime. We therefore employ the Born approximation\cite{Yarimoto:2024uew, Mizuno:2022xxp, Takahashi:2005sxa} and retain only the linear term of the lens potential. The reduced amplification factor is then given by
\begin{eqnarray}
    \eta(w,\bm{y})
    &\simeq&-\frac{w^2}{2\pi}\int d^2x\qty[\psi(\bm{x})-\psi(\bm{y})]e^{\frac{i}{2}w(\bm{x}-\bm{y})^2}\nonumber\\
    &=&2iw\int\frac{d^2q}{(2\pi)^2}\tilde{\kappa}(\bm{q})e^{i\bm{q}\cdot\bm{y}}\frac{e^{-i\frac{q^2}{2w}}-1}{q^2}\,.\label{Eq: eta definition}
\end{eqnarray}
%

%===============================================

\section{Low-frequency expansion of the amplification factor}
\label{Sec: Low-frequency expansion of amplification factor}

In this section, we investigate the low-frequency expansion of the amplification factor and clarify how its expansion coefficients are related to the density profile. To this end, we focus on the low-frequency regime, $w\ll1$ and $wy^2\ll1$ (equivalently $r_{\rm F}\gg R$ and $r_{\rm F}\gg\xi_0\,$). In this regime, the Fresnel scale is much larger than both the characteristic size of the lensing object and the impact parameter. Consequently, the entire lensing object is contained within a Fresnel scale region centered on the point where the line of sight intersects the lens plane.

%===============================================

\subsection{Finite-part prescription}
\label{Sec: Finite part prescription}

To expand $\eta(w,\bm{y})$ in powers of $w\,$, we first split the integration region in Eq. (\ref{Eq: eta definition}) as
\begin{eqnarray}
    \eta(w,\bm{y})=2iw\lim_{\epsilon\to0}\qty(\int_{|\bm{q}|<\epsilon}+\int_{\epsilon<|\bm{q}|})\frac{d^2q}{(2\pi)^2}\tilde{\kappa}(\bm{q})e^{i\bm{q}\cdot\bm{y}}\frac{e^{-i\frac{q^2}{2w}}-1}{q^2}\,.\label{Eq: eta split integration}
\end{eqnarray}
We then examine the contribution from the region $q<\epsilon$ to the integral. If $\tilde{\kappa}(0)$ is finite, corresponding to a finite total mass, the integrand remains finite at $q=0$ and the contribution from $q<\epsilon$ to the integral vanishes in the limit $\epsilon\to0\,$. In Sec. \ref{Sec: Profiles with power-law tails}, we also consider cases in which $\kappa(x)$ exhibits a power-law decay such that $\tilde{\kappa}(0)$ is not finite. Specifically, we restrict our analysis to profiles that decay at infinity, i.e., $\kappa(x)\propto 1/x^s$ with $s>0\,$. In this case, the behavior of $\tilde{\kappa}(q)$ near $q=0$ is $\tilde{\kappa}(q)\propto q^{s-2}\,$. Consequently, the contribution from $q<\epsilon$ to the integral is proportional to $\epsilon^s$ and vanishes in the limit $\epsilon\to0\,$. Based on the above considerations, the contribution from $q<\epsilon$ in Eq. (\ref{Eq: eta split integration}) vanishes. Furthermore, we decompose Eq. (\ref{Eq: eta split integration}) in a form convenient for expanding in powers of $w$ as
\begin{eqnarray}
    \eta(w,\bm{y})=2iw\lim_{\epsilon\to0}\int_{\epsilon<|\bm{q}|}\frac{d^2q}{(2\pi)^2}\frac{\tilde{\kappa}(\bm{q})}{q^2}e^{i\bm{q}\cdot\bm{y}}e^{-i\frac{q^2}{2w}}-2iw\lim_{\epsilon\to0}\int_{\epsilon<|\bm{q}|}\frac{d^2q}{(2\pi)^2}\frac{\tilde{\kappa}(\bm{q})}{q^2}e^{i\bm{q}\cdot\bm{y}}\,.\label{Eq: eta separate}
\end{eqnarray}
Then the nontrivial $w$-dependence is contained only in the first term. However, each term contains a divergence in the limit $\epsilon\to0$ originating from the integration around $q=0\,$. Since $\epsilon$ is introduced in Eq. (\ref{Eq: eta split integration}) as a parameter to split the integration region, $\eta(w,\bm{y})$ is independent of $\epsilon$ and cannot have any divergence arising from the limit $\epsilon\to0$. Therefore, the divergences in the individual terms must cancel each other. To evaluate each term separately, we introduce a symbol $\FP$ to denote the finite part of the integral. For example, the first term in Eq. (\ref{Eq: eta separate}) becomes
\begin{eqnarray}
    &&2iw\lim_{\epsilon\to0}\int_{\epsilon<|\bm{q}|}\frac{d^2q}{(2\pi)^2}\frac{\tilde{\kappa}(\bm{q})}{q^2}e^{i\bm{q}\cdot\bm{y}}e^{-i\frac{q^2}{2w}}\nonumber\\
    &&=2iw\FP_{\epsilon\to0}\int_{\epsilon<|\bm{q}|}\frac{d^2q}{(2\pi)^2}\frac{\tilde{\kappa}(\bm{q})}{q^2}e^{i\bm{q}\cdot\bm{y}}e^{-i\frac{q^2}{2w}}+\lim_{\epsilon\to0}\text{(divergent term in $\epsilon\to0$)}\,.\label{Eq: FP explanation}
\end{eqnarray}
Namely, we define $\FP_{\epsilon\to0}$ as the operation of expanding the expression for small $\epsilon$, removing the divergent terms, and then taking the limit $\epsilon\to0\,$. When removing a divergent term involving powers of $\ln\epsilon\,$, we set the coefficient of $\epsilon$ in the argument of the logarithm to unity\footnote{For example, if a function $f(\epsilon)$ has the asymptotic expansion $f(\epsilon)\simeq a_{-1}/\epsilon+\tilde{a}_0\ln{\epsilon}+a_0+a_1\epsilon+\cdots$ in the limit $\epsilon\to0\,$, then the finite part is defined as $\FP_{\epsilon\to0}f(\epsilon)=a_0\,$.}.
Taking into account that this divergent term cancels the divergence arising from the second term in Eq. (\ref{Eq: eta separate}), we can write Eq. (\ref{Eq: eta separate}) as
\begin{eqnarray}
    \eta(w,\bm{y})=2iw\FP_{\epsilon\to0}\int_{\epsilon<|\bm{q}|}\frac{d^2q}{(2\pi)^2}\frac{\tilde{\kappa}(\bm{q})}{q^2}e^{i\bm{q}\cdot\bm{y}}e^{-i\frac{q^2}{2w}}-2iw\FP_{\epsilon\to0}\int_{\epsilon<|\bm{q}|}\frac{d^2q}{(2\pi)^2}\frac{\tilde{\kappa}(\bm{q})}{q^2}e^{i\bm{q}\cdot\bm{y}}\,.\label{Eq: eta FP intermediate}
\end{eqnarray}

To consider the second term in Eq. (\ref{Eq: eta FP intermediate}), we rewrite the lens potential in Eq. (\ref{Eq: lens potential integral}) as
\begin{eqnarray}
    \psi(\bm{x})=-2\lim_{\epsilon\to0}\int_{\epsilon<|\bm{q}|}\frac{d^2q}{(2\pi)^2}\frac{\tilde{\kappa}(\bm{q})}{q^2}e^{i\bm{q}\cdot\bm{x}}+\mathrm{const.}\label{Eq: lens potential limit}
\end{eqnarray}
Note that since an arbitrary constant can be added to the lens potential, the contribution from $q=0$ is absorbed into this constant term. Similar to Eq. (\ref{Eq: FP explanation}), the term that diverges in the limit $\epsilon\to0$ can also be isolated in Eq. (\ref{Eq: lens potential limit}). For the density profile considered in this paper, this divergent term is a constant independent of $\bm{x}\,$. In particular, for a profile with a finite $\tilde{\kappa}(0)$, the divergent term appearing in Eq. (\ref{Eq: lens potential limit}) is proportional to $\ln \epsilon$ and is a constant independent of $\bm{x}\,$. Even for a profile with a power-law decay and divergent $\tilde{\kappa}(0)\,$, the divergent term is still constant, as discussed in Appendix \ref{App: Asymptotic expansion of the lens potential}. Therefore, the divergence of the integral in Eq. (\ref{Eq: lens potential limit}) can also be absorbed into the constant term. Then, the lens potential can be written as
\begin{eqnarray}
    \psi(\bm{x})=-2\FP_{\epsilon\to0}\int_{\epsilon<|\bm{q}|}\frac{d^2q}{(2\pi)^2}\frac{\tilde{\kappa}(\bm{q})}{q^2}e^{i\bm{q}\cdot\bm{x}}+\psi_*\,,\label{Eq: lens potential FP}
\end{eqnarray}
where $\psi_*$ is an arbitrary constant determined by imposing a boundary condition, for example, at infinity. Substituting Eq. (\ref{Eq: lens potential FP}) into (\ref{Eq: eta FP intermediate}), the reduced amplification factor can be expressed as
\begin{eqnarray}
    \eta(w,\bm{y})=2iw\FP_{\epsilon\to0}\int_{\epsilon<|\bm{q}|}\frac{d^2q}{(2\pi)^2}\frac{\tilde{\kappa}(\bm{q})}{q^2}e^{-i\frac{q^2}{2w}+i\bm{q}\cdot\bm{y}}+iw\qty[\psi(\bm{y})-\psi_*]\,.\label{Eq: eta FP}
\end{eqnarray}
This expression allows us to perform the low-frequency expansion without encountering divergences.

%===============================================

\subsection{Rapidly decaying spherically symmetric profiles}
\label{Sec: Rapidly decaying spherically symmetric profile}

In this subsection, as a first step, we assume that the profile is spherically symmetric and decays faster than any power law for $x\gg1\,$, and investigate the low-frequency expansion of the amplification factor. In Sec. \ref{Sec: Profiles with power-law tails}, we extend the analysis to density profiles with power-law tails. In the first term of Eq. (\ref{Eq: eta FP}), the oscillatory behavior of the exponential factor suppresses contributions from the region $|\bm{q}-w\bm{y}|\gtrsim\sqrt{w}\,$. Since we assume $w\ll1$ and $wy^2\ll1\,$, the saddle point $\bm{q}=w\bm{y}$ is located within the region $q\lesssim\sqrt{w}\,$, and the dominant contribution is restricted to the region $q\lesssim\sqrt{w}\ll1\,$. Then $\tilde{\kappa}(q)$ can be approximated by its Taylor expansion around $q=0\,$, yielding
\begin{eqnarray}
    \tilde{\kappa}(q)=\sum_{m=0}^\infty \frac{(-1)^m}{(2^mm!)^2}M_{2m}q^{2m}\,,\label{Eq: expansion of tildekappa rapidly decay}
\end{eqnarray}
where
\begin{eqnarray}
    M_{2m}=\int d^2x\,x^{2m}\kappa(x)\label{Eq: moment}
\end{eqnarray}
 is the $2m$-th moment of the profile. Eq. (\ref{Eq: expansion of tildekappa rapidly decay}) is given by setting $A_{n}=0$ in Eq. (\ref{Eq: low-momentum expansion}). In this expansion, we have used the assumption that $\kappa(x)$ is spherically symmetric. Furthermore, the finiteness of $M_{2m}$ for arbitrary order is guaranteed by the assumption that $\kappa(x)$ decays faster than any power law. Substituting Eq. (\ref{Eq: expansion of tildekappa rapidly decay}) into Eq. (\ref{Eq: eta FP}) and defining a new function $f_{m}(w,y)$ by\footnote{Although the divergence in $\epsilon\to0$ occurs only for $m=0\,$, we introduce this form to keep the definition uniform for all $m\,$.}
\begin{eqnarray}
    f_{m}(w,y)=2iw\frac{(-1)^m}{(2^mm!)^2}\FP_{\epsilon\to0}\int_{\epsilon<|\bm{q}|}\frac{d^2q}{(2\pi)^2}q^{2m-2}e^{-i\frac{q^2}{2w}+i\bm{q}\cdot\bm{y}}\,,\label{Eq: f_2m definition}
\end{eqnarray}
we obtain
\begin{eqnarray}
    \eta(w,y)=\sum_{m=0}^\infty M_{2m}f_{m}(w,y)+iw\qty[\psi(y)-\psi_*]\,.\label{Eq: eta expansion by f_2m}
\end{eqnarray}

To expand $\eta(w,y)$ in powers of $w\,$, we first expand $f_{m}(w,y)$ in powers of $wy^2\,$. Specifically, for $wy^2\ll1$, and noting that the dominant contribution comes from the region $q\lesssim\sqrt{w}\,$, we can expand the exponential factor in Eq. (\ref{Eq: f_2m definition}) as $e^{i\bm{q}\cdot\bm{y}}=1+i\bm{q\cdot\bm{y}}-(\bm{q}\cdot\bm{y})^2/2+\cdots\,$. This can be verified to be an expansion in powers of $wy^2$ by making the change of variable $\bm{q}=\sqrt{w}\bm{p}$ in Eq. (\ref{Eq: f_2m definition}). According to Appendix \ref{App: Low-frequency expansion formulas}, the low-frequency expansion of $f_{m}(w,y)$ is given by\footnote{Here, $\ln z$ is defined as the principal value for complex $z\,$, i.e., $-\pi<\mathrm{Im}\qty[\ln z]\leq\pi\,$.},
\begin{eqnarray}
    f_{m}(w,y)=
    \begin{dcases}
        iw\frac{\ln(-2iw)-\gamma}{2\pi}+\sum_{k=1}^\infty\frac{1}{\pi kk!}\qty(\frac{iw}{2})^{k+1}y^{2k}&(m=0)\\
        \sum_{k=0}^\infty\frac{(m+k-1)!}{\pi(m!k!)^2}\qty(\frac{iw}{2})^{m+k+1}y^{2k}&(m\geq1)
    \end{dcases}\,,\label{Eq: f_2m expansion}
\end{eqnarray}
where $\gamma$ is Euler's constant. Substituting Eq. (\ref{Eq: f_2m expansion}) into Eq. (\ref{Eq: eta expansion by f_2m}) yields
\begin{eqnarray}
    \eta(w,y)
    &=&i\frac{M_0}{2\pi}w\qty[\ln(-2iw)-\gamma]+\sum_{\substack{m,k\geq0\\(m,k)\neq(0,0)}}M_{2m}y^{2k}\frac{(m+k-1)!}{\pi(m!k!)^2}\qty(\frac{iw}{2})^{m+k+1}\nonumber\\
    &&+\,iw\qty[\psi(y)-\psi_*]\nonumber\\
    &=&i\frac{M_0}{2\pi}w\ln w+\qty{\frac{M_0}{4}+i\qty[\psi(y)-\psi_*+M_0\frac{\ln2-\gamma}{2\pi}]}w\nonumber\\
    &&-\,\frac{M_2+M_0y^2}{4\pi}w^2-i\frac{M_4+4M_2y^2+M_0y^4}{32\pi}w^3+\cdots\,.\label{Eq: eta expansion in w rapid decay}
\end{eqnarray}
This is the low-frequency expansion of $\eta(w,y)$ valid for $w\ll1$ and $wy^2\ll1$ (equivalently $r_{\rm F}\gg R$ and $r_{\rm F}\gg\xi_0\,$). From this expansion, $M_0\,$, $M_2$ and $M_4$ can be extracted successively from the coefficients of $w\ln w\,$, $w^2$ and $w^3\,$, respectively. More generally, it follows that the moments up to $M_{2m}$ can be determined from the expansion coefficients up to $\mathcal{O}(w^{m+1})\,$. However, when $y\gg1$ (equivalently $\xi_0\gg R\,$), $M_0$ becomes dominant in the expansion coefficient at each order in $w\,$. This implies that the low-frequency lensing effect can be well approximated by that of the point-mass lens, when waves pass sufficiently far from the lens. It is worth noting that the expansion formula in Eq. (\ref{Eq: eta expansion in w rapid decay}) has the same form regardless of the choice of $R\,$. Specifically, suppose that we introduce a different scale $R'$ and denote the corresponding dimensionless quantities by $w'\,$, $y'\,$, $M_{2m}'\,$, $\psi'(y')$ and $\psi'_*\,$. Then, the low-frequency expansion in terms of $w'$ is again given by the same form as Eq. (\ref{Eq: eta expansion in w rapid decay}). This follows from the relations $w^{m+1}M_{2m}=w'^{m+1}M_{2m}'\,$, $wy^2=w'y'^2$ and $\,w\psi(y)=w'\psi'(y')\,$, derived from Eqs. (\ref{Eq: profile dimless}), (\ref{Eq: dimless quantity}) and (\ref{Eq: moment}). In particular, the change in the first term of Eq. (\ref{Eq: eta expansion in w rapid decay}) due to $\ln w=\ln w'+2\ln(R/R')$ gives an additional contribution $iM_0w\ln(R/R')/\pi\,$. This contribution is exactly canceled by the change in the term involving $\psi_*\,$. For a rapidly decaying profile, the asymptotic behavior of $\psi(x)$ for $x\gg1$ is given by Eq. (\ref{Eq: lens potential asymptotic}) with $A_{n}=0\,$, namely
\begin{eqnarray}
    \psi(x)\simeq M_0\frac{\ln(x/2)+\gamma}{\pi}+\psi_*\,.\label{Eq: lens potential asymptotic rapid decay}
\end{eqnarray}
Since $\psi'(x')$ has the same asymptotic form, we have $w\psi_*=w'\psi'_*+wM_0\ln(R/R')/\pi\,$. Thus, the term proportional to $\ln(R/R')$ arising from the transformation of $w\ln w$ is canceled by the corresponding change in $-iw\psi_*\,$, leaving Eq. (\ref{Eq: eta expansion in w rapid decay}) invariant under the choice of $R\,$. However, the condition $w\ll1$ for the validity of the expansion relies on the fact that $R$ is chosen as the characteristic length scale of $\Sigma(\xi)\,$. Therefore, when a different scale $R'$ is used, the corresponding condition becomes $w'\ll (R'/R)^2\,$.

\begin{figure}[t]
    \centering
    \begin{minipage}{0.495\columnwidth}
        \centering
        \includegraphics[width=\linewidth, trim={0cm, 0cm, 0.25cm, 0cm}, clip]{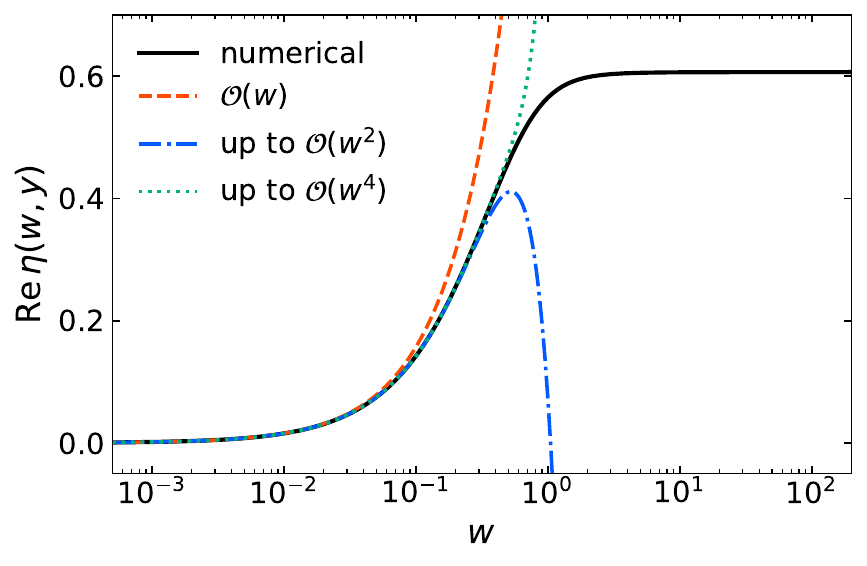}
    \end{minipage}
    \begin{minipage}{0.495\columnwidth}
        \centering
        \includegraphics[width=\linewidth, trim={0.25cm, 0cm, 0cm, 0cm}, clip]{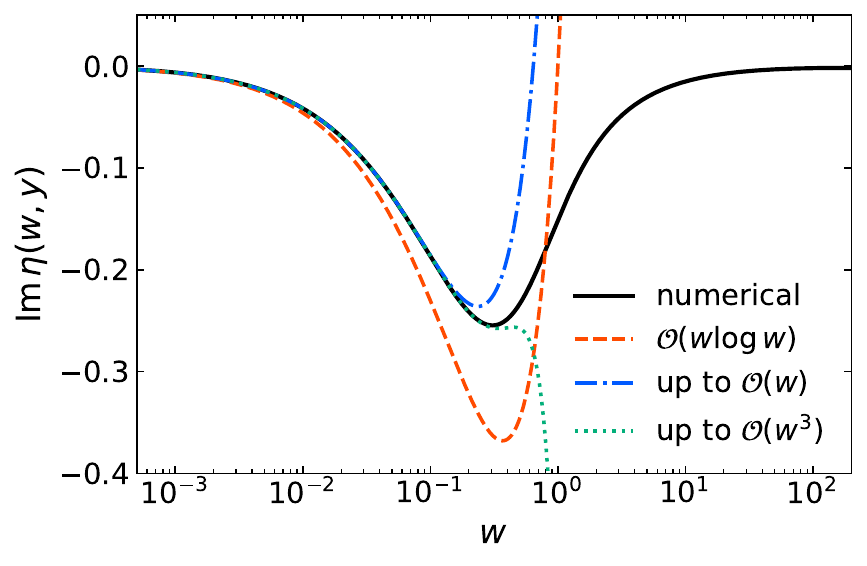}
    \end{minipage}
    \caption{Comparison between $\eta(w,y)$ obtained by numerically evaluating Eq. (\ref{Eq: eta definition}) and its approximations obtained by Eq. (\ref{Eq: eta expansion gauss}) for the Gaussian profile. The profile is normalized as $\kappa(0)=1\,$, and the source position is set to $y=1\,$. The left and right panels show the real and imaginary parts, respectively. The numerical solution (solid line) and three successive approximations obtained from the low-frequency expansion are shown: the lowest-order approximation (dashed line), the next-order approximation (dash-dotted line), and the next higher-order approximation (dotted line). In the real part, terms proportional to $w\,$, $w^2$ and $w^4$ appear in sequence, while in the imaginary part, terms proportional to $w\ln w\,$, $w$ and $w^3$ appear in sequence.}
    \label{Fig: gauss_y_1}
\end{figure}

As an example, we consider the Gaussian density profile of the form
\begin{eqnarray}
    \kappa(x)=\kappa(0)e^{-x^2/2}\,,
\end{eqnarray}
and for simplicity, we set $\kappa(0)=1\,$. Then the moments of this profile are $M_0=2\pi\,$, $M_2=4\pi\,$, $M_4=16\pi$ and $M_6=96\pi\,$. The lens potential is given by\footnote{$\mathrm{E_1}(x)$ is defined by $\int_x^\infty dt\,e^{-t}/t\,$.}
\begin{eqnarray}
    \psi(x)=2\ln x+\mathrm{E_1}\qty(\frac{x^2}{2})\,.
\end{eqnarray}
By comparing the asymptotic expansion of $\psi(x)$ and Eq. (\ref{Eq: lens potential asymptotic rapid decay}), we obtain $\psi_* = 2(\ln2-\gamma)\,$. Substituting the values of $M_{2m}$ and $\psi_*$ into Eq. (\ref{Eq: eta expansion in w rapid decay}), the low-frequency expansion of $\eta(w,y)$ for the Gaussian profile becomes
\begin{eqnarray}
    \eta(w,y)&\simeq&iw\ln w + \qty{\frac{\pi}{2}+i\qty[\psi(y)-\ln 2+\gamma]}w\nonumber\\
    &&-\qty(1+\frac{y^2}{2})w^2-i\qty(\frac{1}{2}+\frac{y^2}{2}+\frac{y^4}{16})w^3+\qty(\frac{1}{3}+\frac{y^2}{2}+\frac{y^4}{8}+\frac{y^6}{144})w^4\,.\label{Eq: eta expansion gauss}
\end{eqnarray}
Fig. \ref{Fig: gauss_y_1} shows $\eta(w,y)$ obtained by numerically evaluating Eq. (\ref{Eq: eta definition}) with $y = 1\,$. We also plot three approximation curves obtained by successively including higher-order terms in Eq. (\ref{Eq: eta expansion gauss}). In the real part, terms proportional to $w\,$, $w^2$ and $w^4$ appear in sequence, while in the imaginary part, terms proportional to $w\ln w\,$, $w$ and $w^3$ appear in sequence. Fig. \ref{Fig: gauss_y_1} shows that these curves approximate $\eta(w,y)$ in the low-frequency regime, $w\ll1$ and $wy^2\ll1$ (which is equivalent to $w\ll1$ here), and it can be confirmed that the accuracy improves as the order of the approximation increases. 

%===============================================

\subsection{Power-law tails and non-analytic frequency dependences}
\label{Sec: Profiles with power-law tails}

In this subsection, we consider how the result obtained in Sec. \ref{Sec: Rapidly decaying spherically symmetric profile}, namely Eq. (\ref{Eq: eta expansion in w rapid decay}) is modified when the profile is spherically symmetric but decays as a power law for $x\gg1\,$. Suppose that, for $x\gg1\,$, $\kappa(x)$ can be expanded in powers of $1/x$ as 
\begin{eqnarray}
    \kappa(x)=\sum_{n=0}^\infty\frac{A_{n}}{x^{s_n}}\,.\label{Eq: profile power law decay}
\end{eqnarray}
Here, $\{s_n\}$ is a sequence of real numbers satisfying $0<s_0<s_1<\cdots\,$, and $A_{n}$ denotes the corresponding expansion coefficients. In this subsection, following Sec. \ref{Sec: Rapidly decaying spherically symmetric profile}, we assume $w\ll1$ and $wy^2\ll1\,$, and similarly expand $\tilde{\kappa}(q)$ with respect to $q\,$. As discussed in Appendix \ref{App: Low-momentum expansion of the density profile}, the moments of the profile can be defined as\footnote{Similar to $\FP_{\epsilon\to0}\,$ we defined $\FP_{\Lambda\to\infty}$ as the operation of expanding the expression for large $\Lambda\,$, removing the divergent terms, and then taking the limit $\Lambda\to\infty\,$.}
\begin{eqnarray}
    M_{2m}=\FP_{\Lambda\to\infty}\int_{|\bm{x}|<\Lambda}d^2x\,x^{2m}\kappa(x)\,,\label{Eq: moment FP}
\end{eqnarray}
and introduce a new function $\chi_s(q)$ for $s>0$ by\footnote{$\mathbb{N}=\{1,2,3,\dots\}$ and $2\mathbb{N}=\{2,4,6,\dots\}$ denotes the set of natural numbers and even numbers, respectively.}
\begin{eqnarray}
    \chi_{s}(q)
    &=&\FP_{\epsilon\to0}\int_{\epsilon<|\bm{x}|}d^2x\frac{e^{-i\bm{q}\cdot\bm{x}}}{x^{s}}\nonumber\\
    &=&
    \begin{dcases}
        \frac{\pi\Gamma(1-s/2)}{\Gamma(s/2)}\qty(\frac{q}{2})^{s-2}&(s\notin2\mathbb{N})\\
        \frac{2\pi(-1)^l}{\qty[(l-1)!]^2}\qty(\frac{q}{2})^{2l-2}\qty(\ln\frac{q}{2}+\gamma-H_{l-1})&(s=2l\,,\,l\in\mathbb{N})
    \end{dcases}\,,
\end{eqnarray}
where $H_{n}$ is the harmonic number defined by $H_0 = 0\,,\,H_{n}=\sum_{k=1}^{n}1/k\quad(n\geq1)\,$. Then $\tilde{\kappa}(q)$ can be expanded for $q\ll1$ as (see Appendix \ref{App: Low-momentum expansion of the density profile})
\begin{eqnarray}
    \tilde{\kappa}(q)=\sum_{m=0}^\infty \frac{(-1)^m}{(2^mm!)^2}M_{2m}q^{2m}+\sum_{n=0}^\infty A_{n}\chi_{s_n}(q)\,.\label{Eq: expansion of tildekappa powerlaw decay}
\end{eqnarray}
Substituting Eq. (\ref{Eq: expansion of tildekappa powerlaw decay}) into Eq. (\ref{Eq: eta FP}) and defining a new function $g_s(w,y)$ for $s>0$ by
\begin{eqnarray}
    g_s(w,y)=2iw\FP_{\epsilon\to0}\int_{\epsilon<|\bm{q}|}\frac{d^2q}{(2\pi)^2}\frac{\chi_s(q)}{q^2}e^{-i\frac{q^2}{2w}+i\bm{q}\cdot\bm{y}}\,,
\end{eqnarray}
we obtain
\begin{eqnarray}
    \eta(w,y)=\sum_{m=0}^\infty M_{2m}f_{m}(w,y)+\sum_{n=0}^\infty A_{n}g_{s_n}(w,y)+iw\qty[\psi(y)-\psi_*]\,.\label{Eq: eta expansion by g_s}
\end{eqnarray}
The expansion of $g_s(w,y)$ in powers of $wy^2$ is presented in Appendix \ref{App: Low-frequency expansion formulas}. As a result, at leading order in $w$ and $wy^2\,$, $g_s(w,y)$ exhibits the following $w$-dependence:
\begin{eqnarray}
     g_s(w,y)\simeq\frac{\Gamma(1-s/2)}{1-s/2}\qty(-\frac{iw}{2})^{s/2}\quad(s\notin2\mathbb{N})\,,\label{Eq: g_s expansion leading}
\end{eqnarray}
\begin{eqnarray}
    g_{2l}(w,y)\simeq
    \begin{dcases}
        -\frac{i}{4}w\ln^2w&(l=1)\\
        -\frac{1}{(l-1)(l-1)!}\qty(\frac{iw}{2})^l\ln w&(l\geq2)
    \end{dcases}\,.\label{Eq: g_2l expansion leading}
\end{eqnarray}
These non-analytic $w$-dependences do not appear when the profile decays faster than a power law and are therefore absent from Eq. (\ref{Eq: eta expansion in w rapid decay}). We note that the low-frequency expansion obtained above reproduces the result derived in Ref. \cite{Choi:2021bkx}\footnote{In Sec. VI of Ref. \cite{Choi:2021bkx}, the leading-order term in the expansion of $F(w,y)$ is derived for $\kappa(x)=(2-k)/(2x^k)$ with $0<k<2\,$. We can verify that this result is reproduced by Eqs. (\ref{Eq: eta expansion by g_s}) and (\ref{Eq: g_s expansion leading}).}.
From the above discussion, we find that, when the profile exhibits a power-law decay as Eq. (\ref{Eq: profile power law decay}), both the moments $M_{2m}$ defined in Eq. (\ref{Eq: moment FP}) and the coefficients of the power-law decay $A_{n}$ can be extracted from the coefficients of the low-frequency expansion of $\eta(w,y)$ in $w\,$.

Note that, we can see from Eq. (\ref{Eq: g_s expansion leading}) that $g_{2l+\delta}(w,y)$ diverges in the limit $\delta\to0\,$. Removing the divergent term with respect to $\delta$ and taking the leading-order term of $w$ gives Eq. (\ref{Eq: g_2l expansion leading}). In Eq. (\ref{Eq: eta expansion by g_s}), the divergence of $g_{2l+\delta}(w,y)$ is canceled by that of $M_{2l-2}$ (for the divergence of $M_{2l-2}\,$, see the discussion below Eq. (\ref{Eq: first term})). Therefore, when Eq. (\ref{Eq: eta expansion by g_s}) is truncated to approximate $\eta(w,y)\,$, if $s_n=2l+\delta\,$, $A_ng_{2l+\delta}(w,y)$ and $M_{2l-2}f_{l-1}(w,y)$ must be retained simultaneously. Otherwise, their divergences remain uncanceled, resulting in an invalid approximation. Furthermore, when Eqs. (\ref{Eq: f_2m expansion}) and (\ref{Eq: g_s expansion}) are truncated to approximate $f_{l-1}(w,y)$ and $g_{2l+\delta}(w,y)\,$, respectively, the corresponding terms in these expansions must also be retained together. Namely, the terms proportional to $w^{l+k}$ in $M_{2l-2}f_{l-1}(w,y)$ and $w^{l+k+\delta/2}$ in $A_{n}g_{2l+\delta}(w,y)$ must be included together, where $k$ is a non-negative integer, since their divergences cancel each other.

\begin{figure}[t]
    \centering
    \begin{minipage}{0.495\columnwidth}
        \centering
        \includegraphics[width=\linewidth, trim={0cm, 0cm, 0.25cm, 0cm}, clip]{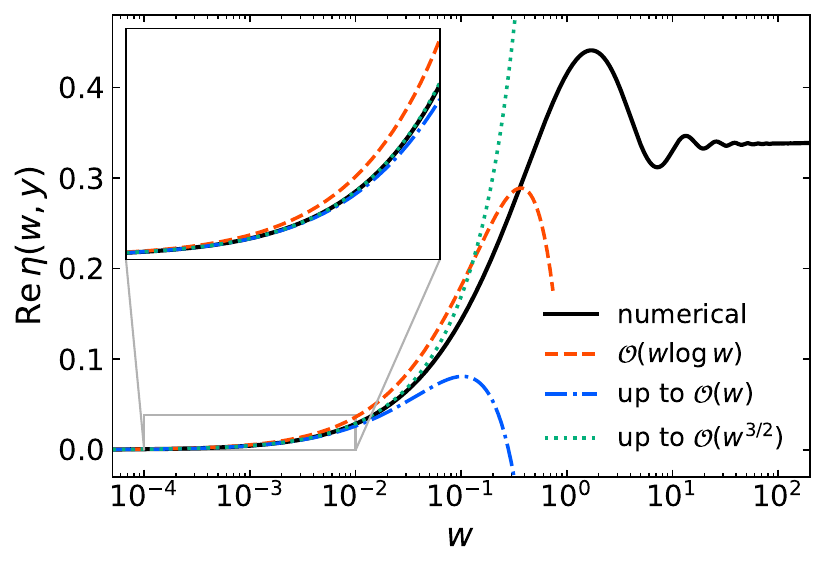}
    \end{minipage}
    \begin{minipage}{0.495\columnwidth}
        \centering
        \includegraphics[width=\linewidth, trim={0.25cm, 0cm, 0cm, 0cm}, clip]{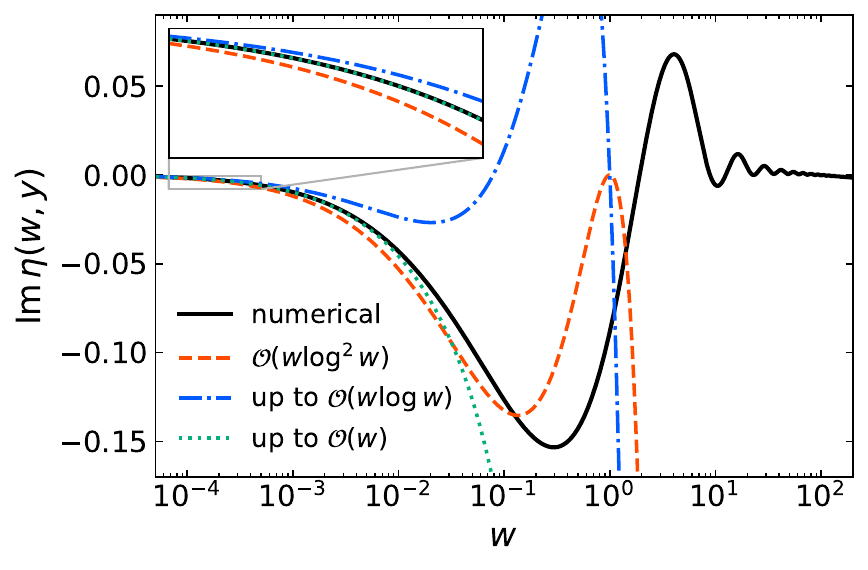}
    \end{minipage}
    \caption{Comparison between $\eta(w,y)$ obtained by numerically evaluating Eq. (\ref{Eq: eta definition}) and its approximations obtained by Eq. (\ref{Eq: eta expansion NFW}) for the NFW profile. The profile is normalized as $\kappa_s=1/2\,$, and the source position is set to $y=1\,$. The left and right panels show the real and imaginary parts, respectively. The numerical solution (solid line) and three successive approximations obtained from the low-frequency expansion are shown: the lowest-order approximation (dashed line), the next-order approximation (dash-dotted line), and the next higher-order approximation (dotted line). In the real part, terms proportional to $w\ln w\,$, $w$ and $w^{3/2}$ appear in sequence, while in the imaginary part, terms proportional to $w\ln^2 w\,$, $w\ln w$ and $w$ appear in sequence.}
    \label{Fig: NFW_y_1}
\end{figure}
As an example, we consider the NFW profile\cite{Navarro:1995iw}. The projected profile is given by\cite{Bartelmann:1996hq}
\begin{eqnarray}
    \kappa(x)=2\kappa_{\rm s}
    \begin{dcases}
        \frac{1}{x^2-1}\qty(1-\frac{2}{\sqrt{1-x^2}}\operatorname{arctanh}\sqrt{\frac{1-x}{1+x}})&(x<1)\\
        \frac{1}{3}&(x=1)\\
        \frac{1}{x^2-1}\qty(1-\frac{2}{\sqrt{x^2-1}}\arctan\sqrt{\frac{x-1}{x+1}})&(x>1)
    \end{dcases}\,,\label{Eq: NFw profile}
\end{eqnarray}
and its Fourier transform is given by\footnote{$\mathrm{Ci}(x)$ and $\mathrm{Si}(x)$ are define by $\mathrm{Ci(x)}=-\int_x^\infty dt \cos t/t$ and $\mathrm{Si(x)}=\int_0^x dt \sin t/t\,$, respectively.}\cite{Cooray:2002dia}
\begin{eqnarray}
    \tilde{\kappa}(q)=4\pi \kappa_{\rm s}\qty{\sin(q)\qty[\frac{\pi}{2}-\mathrm{Si}(q)]-\cos(q)\mathrm{Ci}(q)}\,.
\end{eqnarray}
Here, the dimensionless coordinate $x$ is defined as $x = \xi/r_{\rm s}\,$, where $r_{\rm s}$ is the scale radius. The coefficient $\kappa_{\rm s}$ can be expressed in terms of $r_{\rm s}$ and the characteristic density $\rho_{\rm s}$ as $\kappa_{\rm s}=4\pi Gd_{\rm eff}\rho_{\rm s}r_{\rm s}\,$. For $x\gg1\,$, $\kappa(x)$ can be expanded as $\kappa(x)\simeq 2\kappa_{\rm s}/x^2-\pi\kappa_{\rm s}/x^3\,$, and for simplicity, we set $\kappa_{\rm s}=1/2$ in the following discussion. Comparing this asymptotic behavior with Eq. (\ref{Eq: profile power law decay}), we obtain $A_1=1$ and $A_2=-\pi/2\,$. On the other hand, by comparing the expansion of $\tilde{\kappa}(q)$ for $q\ll1$ and Eq. (\ref{Eq: expansion of tildekappa powerlaw decay}), we obtain $M_0=-2\pi\ln2\,$. The lens potential is given by\cite{Takahashi:2004mc, Meneghetti:2002rn}
\begin{eqnarray}
    \psi(x)=
    \begin{dcases}
        \ln^2\frac{x}{2}-4\operatorname{arctanh}^2\sqrt{\frac{1-x}{1+x}}&(x<1)\\
        \ln^2\frac{x}{2}+4\arctan^2\sqrt{\frac{x-1}{x+1}}&(x\geq1)
    \end{dcases}\,.
\end{eqnarray}
Comparing the asymptotic expansion of $\psi(x)$ and Eq. (\ref{Eq: lens potential asymptotic}), we obtain $\psi_* = \pi^2/4+\gamma^2\,$. Substituting $A_1\,$, $A_2\,$, $M_0$ and $\psi_*$ into Eq. (\ref{Eq: eta expansion by g_s}), the low-frequency expansion of $\eta(w,y)$ for the NFW profile becomes\footnote{The low-frequency behavior of the NFW profile can also be seen in Eq. (51) of Ref. \cite{Choi:2021bkx}. Although the $w\ln^2 w$ term does not appear there, this term arises from the contribution to the integral in Eq. (28) from the region $x\lesssim 1/\sqrt{w}\,$, which is neglected in Ref. \cite{Choi:2021bkx}.}
\begin{eqnarray}
    \eta(w,y)
    &\simeq& M_0f_0(w,y)+A_1g_2(w,y)+A_2g_3(w,y)+iw\qty[\psi(y)-\psi_*]\nonumber\\
    &\simeq&-\frac{i}{4}w\ln^2w-\qty(\frac{\pi}{4}+i\frac{\ln 2+\gamma}{2})w\ln w\nonumber\\
    &&-\qty{\frac{\pi(\ln2+\gamma)}{4}+i\qty[\qty(\frac{\ln2+\gamma}{2})^2+\frac{11\pi^2}{48}-\psi(y)]}w\nonumber\\
    &&+\frac{1+i}{2}\pi^{3/2}w^{3/2}\,.\label{Eq: eta expansion NFW}
\end{eqnarray}
Fig. \ref{Fig: NFW_y_1} shows $\eta(w,y)$ numerically calculated under the Born approximation with $y = 1\,$. We also plot three approximation curves obtained by successively including higher-order terms in Eq. (\ref{Eq: eta expansion NFW}). In the real part, terms proportional to $w\ln w\,$, $w$ and $w^{3/2}$ appear in sequence, while in the imaginary part, terms proportional to $w\ln^2w\,$, $w\ln w$ and $w$ appear in sequence. Fig. \ref{Fig: NFW_y_1} shows that these curves approximate $\eta(w,y)$ in the low-frequency regime, $w\ll1$ and $wy^2\ll1\,$, and the accuracy is seen to improve as higher-order terms are included. Although the improvement in accuracy from the $\mathcal{O}(w\ln^2 w)$ approximation to the $\mathcal{O}(w\ln w)$ approximation is relatively small, the overall trend of increasing accuracy with increasing order is clearly observed. Note that the oscillations observed at $w>1$, whose mathematical description lies beyond the scope of the present formulation, are caused by the interference between waves propagating directly from the source to the observer and waves passing through the lens center, as argued in Ref. \cite{Takahashi:2004mc}. The same reference further points out that the existence of this interference is a consequence of the central divergence of the NFW profile.

%===============================================

\subsection{Effects of nonsphericity}
\label{Sec: Effects of nonsphericity}

So far, we have investigated the low-frequency expansion of the amplification factor for spherically symmetric profiles. We now investigate the frequency dependence induced by a non-spherically symmetric profile. In this subsection, similarly to Sec. \ref{Sec: Rapidly decaying spherically symmetric profile}, we assume that the profile decays faster than any power law and investigate the correction to the low-frequency expansion of the amplification factor arising from the quadrupole moment of the profile. First, the low-momentum expansion of the profile can be written as\footnote{Since the origin of the coordinate is chosen at the center of mass, we have $\int d^2x\,x_i\kappa(\bm{x})=0$.}
\begin{eqnarray}
    \tilde{\kappa}(\bm{q})\simeq M_0-\frac{M_2}{4}q^2-\sum_{i,j}\frac{Q_{ij}}{2}\qty(q_iq_j-\frac{\delta_{ij}}{2}q^2)\,,
\end{eqnarray}
where
\begin{eqnarray}
    Q_{ij}=\int d^2x\qty(x_ix_j-\frac{\delta_{ij}}{2}x^2)\kappa(\bm{x})
\end{eqnarray}
denotes the quadrupole moment. Furthermore, by introducing a new function $h_{ij}(w,\bm{y})$ defined as
\begin{eqnarray}
    h_{ij}(w,\bm{y})=-iw\FP_{\epsilon\to0}\int_{\epsilon<|\bm{q}|}\frac{d^2q}{(2\pi)^2}\qty(\frac{q_iq_j}{q^2}-\frac{\delta_{ij}}{2})e^{-i\frac{q^2}{2w}+i\bm{q}\cdot\bm{y}}\,,
\end{eqnarray}
$\eta(w,\bm{y})$ in the regime $w\ll1$ and $wy^2\ll1$ can be approximated as
\begin{eqnarray}
    \eta(w,\bm{y})\simeq M_0f_0(w,y)+M_2f_2(w,y)+\sum_{i,j}Q_{ij}h_{ij}(w,\bm{y})\,.
\end{eqnarray}
By using the expansion of $f_0(w,y)$ given in Eq. (\ref{Eq: f_2m expansion}), the low-frequency expansion of $h_{ij}(w,\bm{y})$ can be obtained as
\begin{eqnarray}
    h_{ij}(w,\bm{y})
    &=&\frac{1}{2}\qty(\frac{\partial^2}{\partial y_i\partial y_j}-\frac{\delta_{ij}}{2}\nabla_y^2)f_0(w,y)\nonumber\\
    &=&\sum_{k=0}^\infty\frac{2}{\pi(k+2)k!}\qty(\frac{iw}{2})^{k+3}y^{2k}\qty(y_iy_j-\frac{\delta_{ij}}{2}y^2)\,.
\end{eqnarray}
Therefore, the total contributions from the second moments, $M_2$ and $Q_{ij}\,$, is given by
\begin{eqnarray}
    M_2f_2(w,y)+\sum_{i,j}Q_{ij}h_{ij}(w,\bm{y})=\sum_{k=0}^\infty\frac{1}{\pi k!}\qty(\frac{iw}{2})^{k+2}y^{2k}\qty(M_2+\sum_{i,j}\frac{iwy_iy_j}{k+2}Q_{ij})\,.
\end{eqnarray}
This expression shows that the contribution from $Q_{ij}$ is suppressed by $\mathcal{O}(wy^2)$ relative to that from $M_{2}\,$. Therefore, in the low-frequency expansion of the amplification factor, the effect of non-sphericity appears only as a higher-order correction to the spherically symmetric component.

%===============================================

\subsection{Applicability to realistic lens systems}
\label{Sec: Applicability to realistic lens systems}

In this subsection, we investigate up to which order the low-frequency expansion of the amplification factor can be reliably described within the Born approximation. For simplicity, here we assume that the lens profile is spherically symmetric. Furthermore, we restrict our discussion to the case $y\lesssim1\,$. In this case, the conditions $w\ll1$ and $wy^2\ll1$ required for the low-frequency expansion reduce to $w\ll1\,$. We then consider realistic lensing systems that satisfy this low-frequency condition and examine up to which order the Born approximation remains reliable, ensuring the applicability of our low-frequency expansion formula.

Assuming $w\psi\ll1\,$, the leading post-Born correction arises from the term $w^2\psi^2$ when we expand as $e^{-iw\psi}\simeq1-iw\psi-w^2\psi^2/2\,$ in Eq. (\ref{Eq: amplification factor dimless}). Taking into account that $F(w,y)$ has been redefined in Eq. (\ref{Eq: amplification factor redefine}), the Born approximation $\eta_{\rm Born}(w,y)$ and the leading post-Born correction $\Delta\eta_{\rm PB}(w,y)$ are given by
\begin{eqnarray}
    \eta_{\rm Born}(w,y)=-\frac{w^2}{2\pi}\int d^2x\qty[\psi(x)-\psi(y)]e^{\frac{i}{2}w(x-y)^2}\,,\label{Eq: eta Born}
\end{eqnarray}
\begin{eqnarray}
    \Delta\eta_{\rm PB}(w,y)=\frac{iw^3}{4\pi}\int d^2x\qty[\psi(x)-\psi(y)]^2e^{\frac{i}{2}w(x-y)^2}\,.\label{Eq: eta PB}
\end{eqnarray}
Then the amplification factor can be approximated up to the leading post-Born correction as
\begin{eqnarray}
    \eta(w,y)\simeq\eta_{\rm Born}(w,y)+\Delta\eta_{\rm PB}(w,y)\,.
\end{eqnarray}

To estimate the magnitude of $\eta_{\rm Born}(w,y)$ and $\Delta\eta_{\rm PB}(w,y)$, it is useful to first summarize the leading asymptotic behavior of $\psi(x)$ in the regime $x\gg1\,$. According to Eqs. (\ref{Eq: lens potential varphi_s}) and (\ref{Eq: lens potential asymptotic}), up to $\mathcal{O}(1)$ numerical factors\footnote{For $s_1=2+\delta$ with $\delta<0$ and $|\delta|\ll1\,$, the coefficient of $1/x^{s_1-2}$ in Eq. (\ref{Eq: lens potential varphi_s}) can become very large. However, as discussed below Eq. (\ref{Eq: lens potential asymptotic}), asymptotic expansion of $\psi(x)$ can then be approximated by the corresponding expression with $s_1=2\,$. Consequently, the leading behavior of $\psi(x)$ is well approximated by $A_1\ln^2x\,$, and the following discussion can be based on Eq. (\ref{Eq: lens potential leading}), with the omitted numerical coefficients taken to be of order unity.},
it is given by
\begin{eqnarray}
    \psi(x)\sim
    \begin{dcases}
        M_0\ln x&(\text{finite total mass})\\
        A_1\ln^2x&(\text{infinite total mass with $s_1=2$})\\
        \frac{A_1}{x^{s_1-2}}&(\text{infinite total mass with $0<s_1<2$})
    \end{dcases}\,.\label{Eq: lens potential leading}
\end{eqnarray}
Here, the case of finite total mass includes both profiles that decay faster than any power law and profiles with power-law tails of the form $\kappa(x)\simeq A_1/x^{s_1}$ with $s_1>2\,$. Furthermore, the coefficients $M_0$ and $A_1$ are of the same order as the characteristic value of the profile $\kappa_*\,$, namely, 
\begin{eqnarray}
    \kappa_*\sim
    \begin{dcases}
        M_0&(\text{finite total mass})\\
        A_1&(\text{infinite total mass})
    \end{dcases}\,.\label{Eq: kappa_*}
\end{eqnarray}
When the total mass is finite, the definition $M_0=\int d^2x\,\kappa(x)$ together with the fact that the characteristic length scale of $\kappa(x)$ is $x\sim1\,$, implies that $M_0$ can be regarded as the average, and hence a characteristic value, of $\kappa(x)\,$. On the other hand, when the profile has a power-law tail such that the total mass diverges, its characteristic value cannot be defined in terms of its average value. In this case, if the asymptotic behavior $\kappa(x)\simeq A_1/x^{s_1}$ for $x\gg1$ is extrapolated to $x=1\,$, we find $A_1\sim\kappa(1)\,$. Since $x\sim1$ corresponds to the characteristic length scale of the profile, $\kappa(1)$ provides a characteristic value of $\kappa(x)\,$, and hence $A_1\sim\kappa_*\,$. This relation can be explicitly verified for the NFW profile in Eq. (\ref{Eq: NFw profile}).

Then we estimate the integrals in Eqs. (\ref{Eq: eta Born}) and (\ref{Eq: eta PB}) in the regime $w\ll1\,$. The dominant contribution to the integrals comes from the region $|\bm{x}-\bm{y}|\lesssim1/\sqrt{w}$, as the oscillatory exponential factor suppresses contributions from outside this region. Under the assumption $w\ll1$ and $y\lesssim1\,$, this region can be written as $x\lesssim1/\sqrt{w}\,$. Then we approximate $\psi(x)$ by Eq. (\ref{Eq: lens potential leading}) and neglect its logarithmic dependence on $x$ for the purpose of estimating the order of magnitude. Performing the integrals and using Eq. (\ref{Eq: kappa_*}), the leading dependence of $\eta_{\rm Born}(w,y)$ and $\Delta\eta_{\rm PB}(w,y)$ on $w$ can be roughly estimated as\footnote{The post-Born correction of the amplification factor for a point-mass lens has been investigated in Ref. \cite{Yarimoto:2024uew}.}
\begin{eqnarray}
    \eta_{\rm Born}(w,y)\sim w^2\int_{|\bm{x}|<1/\sqrt{w}} d^2x\,\psi(x)\sim\kappa_*
    \begin{dcases}
        w&(\text{finite total mass})\\
        w^{s_1/2}&(\text{infinite total mass})
    \end{dcases}\,,\label{Eq: eta Born leading}
\end{eqnarray}
\begin{eqnarray}
    \Delta\eta_{\rm PB}(w,y)\sim w^3\int_{|\bm{x}|<1/\sqrt{w}} d^2x\,\psi^2(x)\sim\kappa_*
    \begin{dcases}
        w^2&(\text{finite total mass})\\
        w^{s_1}&(\text{infinite total mass})
    \end{dcases}\,.\label{Eq: eta PB leading}
\end{eqnarray}
Here and throughout the following discussion, we neglect the logarithmic dependence on $w$ for the purpose of obtaining an order-of-magnitude estimate. We also neglect the terms arising from $\psi(y)\,$, since they are not important for the present estimate\footnote{The contribution from $\psi(y)$ adds a term proportional to $w\psi(y)$ to $\eta_{\rm Born}(w,y)\,$. In the case of finite total mass, or infinite total mass with $s_1=2\,$, this term has the same $w$-dependence as the leading term in Eq. (\ref{Eq: eta Born leading}). However, using the Poisson equation (\ref{Eq: poisson equation dimless}) and the fact that the characteristic length scale of $\kappa(x)$ is $x\sim1\,$, we find $\psi=\mathcal{O}(\kappa)\,$. Therefore, this contribution only changes a numerical $\mathcal{O}(1)$ factor and does not affect the estimate in Eq. (\ref{Eq: eta Born leading}). On the other hand, in the case of infinite total mass with $0<s_1<2\,$, the term $w\psi(y)$ is sub-leading and can therefore be neglected. A similar argument applies to $\Delta\eta_{\rm PB}(w,y)\,$.}.

Next, we investigate up to which order the terms in the low-frequency expansion of $\eta_{\rm Born}(w,y)$ dominate over $\Delta\eta_{\rm PB}(w,y)\,$. More specifically, we schematically write the low-frequency expansion of $\eta_{\rm Born}(w,y)$ as
\begin{eqnarray}
    \eta_{\rm Born}(w,y)\sim\kappa_*w^{t_0}+\cdots+\kappa_*w^t+\cdots\quad(t_0<t)\,,
\end{eqnarray}
where $t_0$ denotes the order of the leading term given by Eq. (\ref{Eq: eta Born leading}), and $\kappa_*w^t$ represents a higher order term. Here, we assume that all the coefficients in the low-frequency expansion of $\eta_{\rm Born}(w,y)$ are of the same order\footnote{This assumption is valid, at least for the Gaussian and NFW profiles considered in this paper with $y\lesssim1\,$, as can be seen from Eqs. (\ref{Eq: eta expansion gauss}) and (\ref{Eq: eta expansion NFW}).}.
We now derive the condition on $t$ under which the term $\kappa_* w^t$ is larger than $\Delta\eta_{\rm PB}(w,y)\,$. Using Eq. (\ref{Eq: eta PB leading}), this condition becomes
\begin{eqnarray}
    \kappa_*w^t>\kappa_*^2
    \begin{dcases}
        w^2&(\text{finite total mass})\\
        w^{s_1}&(\text{infinite total mass})
    \end{dcases}\,,
\end{eqnarray}
or equivalently for $w<1\,$,
\begin{equation}
    t<t_*=\frac{\ln\kappa_*}{\ln w}+
    \begin{dcases}
        2&(\text{finite total mass})\\
        s_1&(\text{infinite total mass})
    \end{dcases}\,,\label{Eq: threshold order}
\end{equation}
where we defined the threshold order as $t_*\,$, which depends on the frequency. To make this criterion more concrete, let us consider a given frequency $w\ll1\,$, and suppose that $t_0<t_1<t_*<t_2$ and that the low-frequency expansion of $\eta_{\rm Born}(w,y)$ takes the form
\begin{eqnarray}
    \eta_{\rm Born}(w,y)\sim\kappa_*w^{t_0}+\cdots+\kappa_*w^{t_1}+\kappa_*w^{t_2}+\cdots\,.
\end{eqnarray}
Since $\kappa_*w^{t_1}>\Delta\eta_{\rm PB}(w,y)>\kappa_*w^{t_2}\,$, the amplification factor approximated up to the leading post-Born order is given schematically by
\begin{eqnarray}
    \eta(w,y)\sim\kappa_*w^{t_0}+\cdots+\kappa_*w^{t_1}+\Delta\eta_{\rm PB}(w,y)\,.
\end{eqnarray}
As can be seen from this equation, the low-frequency expansion can be reliably described within the Born approximation up to order $w^{t_1}\,$.

We now consider a realistic lensing system satisfying $w\ll1$ and examine to which order our low-frequency expansion formula within the Born approximation can be used. As such a lensing system, we consider a dark matter subhalo with a virial mass of $10^6 M_{\odot}\,$. Assuming an NFW density profile, such a subhalo has a scale radius $r_{\rm s}\simeq50\,\mathrm{pc}$ and a characteristic surface density $\Sigma_{\rm s}=\rho_{\rm s}r_{\rm s}\simeq10^7 M_{\odot}/\mathrm{kpc}^2\,$\footnote{These rough estimates were obtained using the mass-concentration relation given in Ref. \cite{Dutton:2014xda}}. We also take the parameters of the gravitational-wave source to be $d_{\rm eff}=10\,\mathrm{Gpc}$ and $f=1\,\mathrm{mHz}\,$. For this configuration, using $w=(r_{\rm s}/r_{\rm F})^2$ together with Eq. (\ref{Eq: Fresnel scale}), we obtain $w\simeq0.2\,$, and thus the condition $w\ll1$ is satisfied. Nondimensionalizing $\Sigma_{\rm s}$ according to Eq. (\ref{Eq: profile dimless}) gives $\kappa_*\simeq0.06\,$. Substituting these values and $s_1=2$\footnote{The NFW profile decays for $x\gg1$ as $\kappa\propto1/x^2\,$, as can be seen from Eq. (\ref{Eq: NFw profile}).}
into Eq. (\ref{Eq: threshold order}), we obtain $t_*\simeq3.5\,$. Therefore, in this case, the low-frequency expansion of $\eta(w,y)$ in the Born approximation can be reliably used up to terms of order $w^3\,$. At order $w^{3.5}\,$, the post-Born correction becomes comparable and can no longer be neglected.

%===============================================

\section{Conclusion}
\label{Sec: Conclusion}

In this paper, we investigated the low-frequency behavior of the amplification factor in gravitational lensing and explored how information about the density profile of the lensing object is encoded. Under the Born approximation, we derived a general low-frequency expansion of the amplification factor for a broad class of density profiles. For spherically symmetric profiles that decay faster than any power law, we showed that each coefficient in the expansion is determined by a finite number of moments of the density profile, establishing a correspondence between the low-frequency behavior of the amplification factor and the internal structure of the lensing object. We further found that density profiles with power-law tails generate additional non-analytic frequency dependences, including fractional-power and logarithmic terms\footnote{Note that even in the case of rapidly decaying profiles, a logarithmic dependence appears in the leading term of the low-frequency expansion.},
which reflect the asymptotic behavior of the density profile. Therefore, the low-frequency expansion provides information not only about the internal structure of the lensing object but also about its asymptotic density distribution. We also examined the effects of nonsphericity and showed that deviations from spherical symmetry appear only as higher-order corrections in the low-frequency regime. Finally, we investigated the validity of the Born approximation in the low-frequency expansion. We derive a criterion for the maximum order of the low-frequency expansion up to which the Born approximation remains dominant over post-Born corrections.

Taken together, these results demonstrate that the low-frequency expansion of the amplification factor provides a systematic framework for extracting information about density profiles. In particular, the expansion coefficients are directly related to the moments of the density profile and to the coefficients characterizing its power-law tail, thereby providing a model-independent parametrization of wave-optical lensing signals. This framework will be useful for inferring density profiles of lensing objects from future gravitational wave observations. An important direction for future work is to investigate to what extent the coefficients of the low-frequency expansion can be measured in realistic gravitational wave observations and how accurately the underlying density profile can be reconstructed from them.

%===============================================

\section*{Acknowledgements}
This work was supported by JSPS KAKENHI Grant Number 24KJ1094 (ST) and Grant Number 26K07083 (TS).

%===============================================

\appendix

%===============================================

\section{Low-momentum expansion of the density profile}
\label{App: Low-momentum expansion of the density profile}

In this section, we derive the low-momentum expansion of the profile. We assume that the profile is spherically symmetric and admits the following asymptotic expansion in powers of $1/x$ for $x\gg1\,$:
\begin{eqnarray}
    \kappa(x)=\sum_{n=0}^\infty\frac{A_{n}}{x^{s_n}}\,,\label{Eq: asymptotic expansion of profile}
\end{eqnarray}
where $\{s_n\}$ is a sequence of real numbers satisfying $0<s_0<s_1<\cdots\,$. We first split the integration range in the Fourier transform of $\kappa(x)$ as
\begin{eqnarray}
    \tilde{\kappa}(q)=\lim_{\epsilon\to0}\qty(\int_{|\bm{x}|<\epsilon}+\int_{\epsilon<|\bm{x}|})d^2x\,\kappa(x)e^{-i\bm{q}\cdot\bm{x}}\,.\label{Eq: tildekappa split}
\end{eqnarray}
Assuming that the mass enclosed within an infinitesimal region around the origin vanishes,
\begin{equation}
    \lim_{\epsilon\to0}\int_{|\bm{x}|<\epsilon}d^2x\,\kappa(x)=0\,,\label{Eq: assumption origin}
\end{equation}
the contribution from $x<\epsilon$ in Eq. (\ref{Eq: tildekappa split}) vanishes. Furthermore, since $\epsilon$ is merely to split the integration range and $\tilde{\kappa}(q)$ is independent of $\epsilon\,$, $\tilde{\kappa}(q)$ does not contain any divergence associated with the limit $\epsilon\to0\,$. Therefore, $\lim_{\epsilon\to0}$ can be replaced by $\FP_{\epsilon\to0}$ in Eq. (\ref{Eq: tildekappa split}), yielding
\begin{eqnarray}
    \tilde{\kappa}(q)=\FP_{\epsilon\to0}\int_{\epsilon<|\bm{x}|}d^2x\,\kappa(x)e^{-i\bm{q}\cdot\bm{x}}\,.\label{Eq: tildekappa FP}
\end{eqnarray}
Then we isolate the asymptotic behavior of $\kappa(x)$ by decomposing the above equation as
\begin{eqnarray}
    \tilde{\kappa}(q)=\FP_{\epsilon\to0}\int_{\epsilon<|\bm{x}|}d^2x\qty[\kappa(x)-\sum_{n=0}^\infty\frac{A_{n}}{x^{s_n}}]e^{-i\bm{q}\cdot\bm{x}}+\sum_{n=0}^\infty A_{n}\FP_{\epsilon\to0}\int_{\epsilon<|\bm{x}|}d^2x\frac{e^{-i\bm{q}\cdot\bm{x}}}{x^{s_n}}\,.\label{Eq: tildekappa FP decomposed}
\end{eqnarray}

We now examine the first term in Eq. (\ref{Eq: tildekappa FP decomposed}). Since the integrand decays sufficiently rapidly for $x\gtrsim1\,$, only the region $x\lesssim1$ contributes to the integral. Assuming $q\ll1\,$, we can expand the exponential term $e^{-i\bm{q}\cdot\bm{x}}$ to obtain
\begin{eqnarray}
    &&\FP_{\epsilon\to0}\int_{\epsilon<|\bm{x}|}d^2x\qty[\kappa(x)-\sum_{n=0}^\infty\frac{A_{n}}{x^{s_n}}]e^{-i\bm{q}\cdot\bm{x}}\nonumber\\
    &&=\sum_{m=0}^\infty\frac{(-q^2)^m}{(2^mm!)^2}\FP_{\epsilon\to0}\int_{\epsilon<|\bm{x}|}d^2x\,x^{2m}\qty[\kappa(x)-\sum_{n=0}^\infty\frac{A_{n}}{x^{s_n}}]\,,\label{Eq: tildekappa FP first term}
\end{eqnarray}
where we have used
\begin{eqnarray}
    \int_0^{2\pi}\frac{d\theta}{2\pi}\cos^{2m}\theta=\frac{(2m)!}{(2^mm!)^2}\,.\label{Eq: theta integration}
\end{eqnarray}
Since integrand of the integral appearing on the right-hand side of Eq. (\ref{Eq: tildekappa FP first term}) decays faster than any power law for $x\gg1\,$, the integral does not exhibit any divergence associated with the upper limit of the integration. Given this, following the same procedure as in the derivation from Eqs. (\ref{Eq: tildekappa split}) to (\ref{Eq: tildekappa FP}), the integral can be written with $\FP_{\Lambda\to\infty}$ as
\begin{eqnarray}
    \FP_{\epsilon\to0}\int_{\epsilon<|\bm{x}|}d^2x\,x^{2m}\qty[\kappa(x)-\sum_{n=0}^\infty\frac{A_{n}}{x^{s_n}}]&=&\FP_{\epsilon\to0,\Lambda\to\infty}\int_{\epsilon<|\bm{x}|<\Lambda}d^2x\,x^{2m}\kappa(x)\nonumber\\
    &&-\sum_{n=0}^\infty A_n\FP_{\epsilon\to0,\Lambda\to\infty}\int_{\epsilon<|\bm{x}|<\Lambda}d^2x\,x^{2m-s_n}\,.
\end{eqnarray}
The second term on the right-hand side in the above equation vanishes after performing the integral and taking the finite part. Since the first term is finite in the limit $\epsilon\to0\,$, as follows from Eq. (\ref{Eq: assumption origin}), the finite-part prescription with respect to $\epsilon$ can be omitted. Thus, we obtain
\begin{eqnarray}
    \FP_{\epsilon\to0}\int_{\epsilon<|\bm{x}|}d^2x\,x^{2m}\qty[\kappa(x)-\sum_{n=0}^\infty\frac{A_{n}}{x^{s_n}}]=\FP_{\Lambda\to\infty}\int_{|\bm{x}|<\Lambda}d^2x\,x^{2m}\kappa(x)\,.
\end{eqnarray}
Substituting the above equation into Eq. (\ref{Eq: tildekappa FP first term}), and defining the moments of the profile by
\begin{eqnarray}
    M_{2m}=\FP_{\Lambda\to\infty}\int_{|\bm{x}|<\Lambda}d^2x\,x^{2m}\kappa(x)\,,\label{Eq: moment FP appendix}
\end{eqnarray}
we obtain
\begin{eqnarray}
    \FP_{\epsilon\to0}\int_{\epsilon<|\bm{x}|}d^2x\qty[\kappa(x)-\sum_{n=0}^\infty\frac{A_{n}}{x^{s_n}}]e^{-i\bm{q}\cdot\bm{x}}=\sum_{m=0}^\infty\frac{(-1)^m}{(2^mm!)^2}M_{2m}q^{2m}\,.\label{Eq: first term}
\end{eqnarray}
For later use, here we examine the divergence of $M_{2m}\,$. Suppose that the asymptotic expansion of $\kappa(x)$ in Eq. (\ref{Eq: asymptotic expansion of profile}) contains a term with $s_n=2l+\delta\,$, where $l\in\mathbb{N}$ and $|\delta|\ll1\,$. The contribution of $A_n/x^{s_n}$ to $M_{2l-2}$ from the region $1\ll a<x<\Lambda$ is
\begin{eqnarray}
    \FP_{\Lambda\to\infty}\int_{a<|\bm{x}|<\Lambda}d^2x\,x^{2l-2}\frac{A_{n}}{x^{2l+\delta}}=
    \begin{dcases}
        -2\pi A_n\ln a&(\delta=0)\\
        \frac{2\pi A_{n}}{\delta a^\delta}\simeq2\pi A_n\qty(\frac{1}{\delta}-\ln a) &(\delta\neq0)
    \end{dcases}\,,
\end{eqnarray}
where $a\gg1$ is an arbitrary constant. The above equation shows that, for $s_n=2l+\delta\,$, $M_{2l-2}$ has a divergence proportional to $1/\delta$ in the limit $\delta\to0\,$, whose coefficient is independent of the choice of $a\,$. By subtracting this divergent part, one recovers $M_{2l-2}$ for $s_n=2l\,$. In other words, the definition of the moments in Eq. (\ref{Eq: moment FP appendix}) is discontinuous at $s_n=2l$ in this sense. As will be seen below, the divergence discussed here, which arises from the first term in Eq. (\ref{Eq: tildekappa FP decomposed}), is canceled by the divergence arising from the second term in Eq. (\ref{Eq: tildekappa FP decomposed}).

For the second term of Eq. (\ref{Eq: tildekappa FP decomposed}), we introduce $\chi_s(q)$ for $s>0$ as
\begin{eqnarray}
    \chi_s(q)=\FP_{\epsilon\to0}\int_{\epsilon<|\bm{x}|}d^2x\frac{e^{-i\bm{q}\cdot\bm{x}}}{x^{s}}\,.\label{Eq: chi definition}
\end{eqnarray}
Then Eq. (\ref{Eq: tildekappa FP decomposed}) becomes
\begin{eqnarray}
    \tilde{\kappa}(q)=\sum_{m=0}^\infty\frac{(-1)^m}{(2^mm!)^2}M_{2m}q^{2m}+\sum_{n=0}^\infty A_{n}\chi_{s_n}(q)\,.\label{Eq: low-momentum expansion}
\end{eqnarray}
In what follows, we derive the explicit expression for $\chi_s(q)\,$. For $s\notin2\mathbb{N}\,$, the integral in Eq. (\ref{Eq: chi definition}) can be expanded by infinitesimal $\epsilon$ as
\begin{eqnarray}
    \int_{\epsilon<|\bm{x}|}d^2x\frac{e^{-i\bm{q}\cdot\bm{x}}}{x^{s}}=\frac{\pi\Gamma(1-s/2)}{\Gamma(s/2)}\qty(\frac{q}{2})^{s-2}+\sum_{n=0}^\infty\frac{2\pi(-q^2)^n}{(2^nn!)^2}\frac{\epsilon^{2n+2-s}}{s-2n-2}\,.\label{Eq: profile integral}
\end{eqnarray}
Taking the limit $s\to2l$ for $l\in\mathbb{N}$ in the above equation yields
\begin{eqnarray}
    \int_{\epsilon<|\bm{x}|}d^2x\frac{e^{-i\bm{q}\cdot\bm{x}}}{x^{2l}}&=&\frac{2\pi(-1)^l}{\qty[(l-1)!]^2}\qty(\frac{q}{2})^{2l-2}\qty(\ln\frac{q}{2}+\gamma-H_{l-1})\nonumber\\
    &&-\,\frac{2\pi(-q^2)^{l-1}}{\qty[2^{l-1}(l-1)!]^2}\ln\epsilon+\sum_{n\geq0,n\neq l-1}\frac{\pi(-q^2)^n}{(2^nn!)^2}\frac{\epsilon^{2n+2-2l}}{l-n-1}\,,\label{Eq: profile integral log}
\end{eqnarray}
where $H_{n}$ is the harmonic number defined by $H_0 = 0\,,\,H_{n}=\sum_{k=1}^{n}1/k\quad(n\geq1)\,$. Finally, by taking the finite part in the limit $\epsilon\to0$ of Eqs. (\ref{Eq: profile integral}) and (\ref{Eq: profile integral log}), we obtain
\begin{eqnarray}
    \chi_s(q)=
    \begin{dcases}
        \frac{\pi\Gamma(1-s/2)}{\Gamma(s/2)}\qty(\frac{q}{2})^{s-2}&(s\notin2\mathbb{N})\\
        \frac{2\pi(-1)^l}{\qty[(l-1)!]^2}\qty(\frac{q}{2})^{2l-2}\qty(\ln\frac{q}{2}+\gamma-H_{l-1})&(s=2l)
    \end{dcases}\,.\label{Eq: chi expression}
\end{eqnarray}
Note that $\chi_{2l+\delta}(q)$ diverges in the limit $\delta\to0\,$, and subtracting the term that diverges as $\delta\to0$ gives $\chi_{2l}(q)\,$. However, in Eq. (\ref{Eq: low-momentum expansion}), the divergence of $\chi_{2l+\delta}(q)$ is canceled by the divergence of $M_{2l-2}$ discussed below Eq. (\ref{Eq: first term}). Therefore, although $\chi_{s_n}(q)$ and $M_{2l-2}$ are each discontinuous at $s_n=2l\,$, their combination in Eq. (\ref{Eq: low-momentum expansion}) is continuous at $s_n=2l\,$. Based on the above discussion, when Eq. (\ref{Eq: low-momentum expansion}) is truncated to approximate $\tilde{\kappa}(q)\,$, the terms involving $\chi_{2l+\delta}(q)$ and $M_{2l-2}\,$ must always be retained together. Otherwise, their divergences do not cancel, leading to an invalid approximation.

%===============================================

\section{Asymptotic expansion of the lens potential}
\label{App: Asymptotic expansion of the lens potential}

Here we examine the asymptotic expansion of the lens potential for profiles with power-law tails. As in Appendix \ref{App: Low-momentum expansion of the density profile}, we assume that the profile is spherically symmetric and decays according to the power law given in Eq. (\ref{Eq: asymptotic expansion of profile}). First, we express the potential as
\begin{eqnarray}
    \psi(x)=-2\FP_{\epsilon\to0}\int_{\epsilon<|\bm{q}|}\frac{d^2q}{(2\pi)^2}\frac{\tilde{\kappa}(q)}{q^2}e^{i\bm{q}\cdot\bm{x}}+\psi_*\,,\label{Eq: lens potential definition App}
\end{eqnarray}
where $\psi_*$ is an arbitrary constant. Since the dominant contribution comes from the region $q\ll1$ for $x\gg1\,$, $\tilde{\kappa}(q)$ can be expanded as in Eq. (\ref{Eq: low-momentum expansion}), and we obtain
\begin{eqnarray}
    \psi(x)&=&-2\sum_{m=0}^\infty\frac{(-1)^{m}}{(2^mm!)^2}M_{2m}\FP_{\epsilon\to0}\int_{\epsilon<|\bm{q}|}\frac{d^2q}{(2\pi)^2}q^{2m-2}e^{i\bm{q}\cdot\bm{x}}\nonumber\\
    &&-2\sum_{n=0}^\infty A_{n}\FP_{\epsilon\to0}\int_{\epsilon<|\bm{q}|}\frac{d^2q}{(2\pi)^2}\frac{\chi_{s_n}(q)}{q^2}e^{i\bm{q}\cdot\bm{x}}+\psi_*\,.\label{Eq: lens potential expansion}
\end{eqnarray}
To evaluate the integrals appearing in the above equation, we use the following formulas:
\begin{equation}
    \int_{\epsilon<|\bm{q}|}\frac{d^2q}{(2\pi)^2}q^{t-2}e^{i\bm{q}\cdot\bm{x}}=
    \begin{dcases}
        -\frac{\ln(x/2)+\gamma}{2\pi}-\frac{\ln\epsilon}{2\pi}+\mathcal{O}(\epsilon^2)&(t=0)\\
        \frac{\Gamma(t/2)}{4\pi\Gamma(1-t/2)}\qty(\frac{2}{x})^t-\frac{\epsilon^t}{2\pi t}+\mathcal{O}(\epsilon^{t+2})&(t>-2\,,\,t\neq0)
    \end{dcases}\,,\label{Eq: divergent integral q}
\end{equation}
\begin{equation}
    \int_{\epsilon<|\bm{q}|}\frac{d^2q}{(2\pi)^2}q^{2n-2}\ln q\,e^{i\bm{q}\cdot\bm{x}}=
    \begin{dcases}
        \frac{\qty[\ln(x/2)+\gamma]^2}{4\pi}+\mathcal{O}(\ln^2\epsilon)&(n=0)\\
        \frac{(-1)^{n}[(n-1)!]^2}{8\pi}\qty(\frac{2}{x})^{2n}+\mathcal{O}(\epsilon^{2n}\ln\epsilon)&(n\geq1)
    \end{dcases}\,.\label{Eq: divergent integral logq}
\end{equation}
The second equation is obtained by differentiating the first equation with respect to $t$ and then taking the limit $t\to2n\,$ for $n\in\{0,1,2,\dots\}\,$. Note that, the terms in Eq. (\ref{Eq: divergent integral q}) that diverge in the limit $\epsilon\to0$ are independent of $x\,$, and hence the divergent terms in Eq. (\ref{Eq: divergent integral logq}) are also independent of $x\,$. Therefore, the term removed by the finite-part prescription in Eq. (\ref{Eq: lens potential definition App}) is a constant, which justifies the transition from Eq. (\ref{Eq: lens potential limit}) to Eq. (\ref{Eq: lens potential FP}). Using Eq. (\ref{Eq: divergent integral q}), the first term in Eq. (\ref{Eq: lens potential expansion}) is nonzero only for $m=0\,$. For the second term in Eq. (\ref{Eq: lens potential expansion}), Eqs. (\ref{Eq: chi expression}), (\ref{Eq: divergent integral q}) and (\ref{Eq: divergent integral logq}) allow us to rewrite it as
\begin{eqnarray}
    -2\FP_{\epsilon\to0}\int_{\epsilon<|\bm{q}|}\frac{d^2q}{(2\pi)^2}\frac{\chi_{s_n}(q)}{q^2}e^{i\bm{q}\cdot\bm{x}}=\varphi_{s_n}(x)\,.
\end{eqnarray}
Here, $\varphi_{s}(x)$ is defined by
\begin{eqnarray}
    \varphi_{s}(x)=
    \begin{dcases}
        \ln^2x-(\ln2 -\gamma)^2&(s=2)\\
        \frac{2}{(s-2)^2}\frac{1}{x^{s-2}}&(s>0\,,\,s\neq2)
    \end{dcases}\,,\label{Eq: lens potential varphi_s}
\end{eqnarray}
which represents the asymptotic behavior of the potential corresponding to the decay of $\kappa(x)$ proportional to $1/x^s\,$. Finally, we obtain the asymptotic expansion of $\psi(x)$ for $x\gg1$ as
\begin{eqnarray}
    \psi(x)= M_0\frac{\ln(x/2)+\gamma}{\pi}+\sum_{n=0}^\infty A_{n}\varphi_{s_n}(x)+\psi_*\,.\label{Eq: lens potential asymptotic}
\end{eqnarray}
Note that, if the asymptotic expansion of $\kappa(x)$ in Eq. (\ref{Eq: asymptotic expansion of profile}) contains a term with $s_n=2+\delta\,$, $\varphi_{2+\delta}(x)$ appears in Eq. (\ref{Eq: lens potential asymptotic}) and diverges in the limit $\delta\to0$ as $\varphi_{2+\delta}(x)\simeq 2/\delta^2-2\ln x/\delta+\ln^2x\,$. In this case, as discussed below Eq. (\ref{Eq: first term}), $M_0$ also has a divergence proportional to $1/\delta\,$. In Eq. (\ref{Eq: lens potential asymptotic}), the divergence proportional to $\ln x/\delta$ arising from $\varphi_{2+\delta}(x)$ is canceled by that arising from $M_0\ln x\,$. The $x$-independent divergent terms, namely the term proportional to $1/\delta^2$ arising from $\varphi_{2+\delta}(x)$ and the term proportional to $1/\delta$ arising from the first term of Eq. (\ref{Eq: lens potential asymptotic}), can be absorbed into the arbitrary constant $\psi_*\,$. Consequently, after an appropriate redefinition of $\psi_*\,$, the resulting asymptotic expansion of $\psi(x)$ reduces to the form obtained for $s_n=2\,$, involving $\varphi_2(x)$ defined in Eq. (\ref{Eq: lens potential varphi_s}). This shows that, when the asymptotic expansion of $\kappa(x)$ contains a term with $s_n=2+\delta$ for small $\delta\,$, the asymptotic expansion of $\psi(x)$ can be approximated by that obtained by setting $s_n=2\,$.

%===============================================

\section{Evaluation of Fresnel-type integrals}
\label{App: Evaluation of Fresnel-type integrals}

Here we evaluate Fresnel-type integrals used throughout this paper. For $t>-2$ and $t\neq0\,$,
\begin{eqnarray}
    \int_{\epsilon<|\bm{q}|}\frac{d^2q}{(2\pi)^2}q^{t-2}e^{-i\frac{q^2}{2w}}=\frac{\Gamma(t/2)(-2iw)^{t/2}}{4\pi}-\frac{\epsilon^t}{2\pi t}+\mathcal{O}(\epsilon^{t+2})
\end{eqnarray}
holds in the limit $\epsilon\to0\,$. By taking $t\to0$ in the above equation,
\begin{eqnarray}
    \int_{\epsilon<|\bm{q}|}\frac{d^2q}{(2\pi)^2}\frac{e^{-i\frac{q^2}{2w}}}{q^2}=\frac{\ln(-2iw)-\gamma}{4\pi}-\frac{\ln\epsilon}{2\pi}+\mathcal{O}(\epsilon^2)\,.
\end{eqnarray}
Then we obtain
\begin{eqnarray}
    \FP_{\epsilon\to0}\int_{\epsilon<|\bm{q}|}\frac{d^2q}{(2\pi)^2}q^{t-2}e^{-i\frac{q^2}{2w}}=
    \begin{dcases}
        \frac{\ln(-2iw)-\gamma}{4\pi}&(t=0)\\
        \frac{\Gamma(t/2)(-2iw)^{t/2}}{4\pi}&(t>-2\,,\,t\neq0)
    \end{dcases}\,.\label{Eq: fresnel type integral}
\end{eqnarray}

We also evaluate another integral. For integers $n\geq1\,$,
\begin{eqnarray}
    \FP_{\epsilon\to0}\int_{\epsilon<|\bm{q}|}\frac{d^2q}{(2\pi)^2}q^{2n-2}\ln q\,e^{-i\frac{q^2}{2w}}
    &=&\FP_{\epsilon\to0}\frac{d}{dt}\int_{\epsilon<|\bm{q}|}\frac{d^2q}{(2\pi)^2}q^{t-2}\,e^{-i\frac{q^2}{2w}}\bigg|_{t=2n}\nonumber\\
    &=&\frac{(n-1)!(-2iw)^n\qty[\ln(-2iw)-\gamma+H_{n-1}]}{8\pi}\,.
\end{eqnarray}
For $n=0\,$,
\begin{eqnarray}
    \FP_{\epsilon\to0}\int_{\epsilon<|\bm{q}|}\frac{d^2q}{(2\pi)^2}\frac{\ln q}{q^2}\,e^{-i\frac{q^2}{2w}}
    &=&\FP_{\epsilon\to0}\lim_{t\to0}\frac{d}{dt}\int_{\epsilon<|\bm{q}|}\frac{d^2q}{(2\pi)^2}q^{t-2}\,e^{-i\frac{q^2}{2w}}\nonumber\\
    &=&\FP_{\epsilon\to0}\qty{\frac{\qty[\ln(-2iw)-\gamma]^2}{16\pi}+\frac{\pi}{96}-\frac{\ln^2\epsilon}{2\pi}+\mathcal{O}(\epsilon^2\ln\epsilon)}\nonumber\\
    &=&\frac{\qty[\ln(-2iw)-\gamma]^2}{16\pi}+\frac{\pi}{96}
\end{eqnarray}
In summary, we obtain
\begin{eqnarray}
    \FP_{\epsilon\to0}\int_{\epsilon<|\bm{q}|}\frac{d^2q}{(2\pi)^2}q^{2n-2}\ln q\,e^{-i\frac{q^2}{2w}}=
    \begin{dcases}
        \frac{\qty[\ln(-2iw)-\gamma]^2}{16\pi}+\frac{\pi}{96}&(n=0)\\
        \frac{(n-1)!(-2iw)^n\qty[\ln(-2iw)-\gamma+H_{n-1}]}{8\pi}&(n\geq1)
    \end{dcases}\,.\nonumber\\
    \label{Eq: fresnel type integral log}
\end{eqnarray}
%

%===============================================

\section{Low-frequency expansion formulas}
\label{App: Low-frequency expansion formulas}

In this section, we derive the low-frequency expansions of $f_{m}(w,y)$ and $g_s(w,y)$ in the regime $wy^2\ll1\,$. First, $f_{m}(w,y)$ is defined for integers $m\geq0$ by
\begin{eqnarray}
    f_{m}(w,y)=2iw\frac{(-1)^m}{(2^mm!)^2}\FP_{\epsilon\to0}\int_{\epsilon<|\bm{q}|}\frac{d^2q}{(2\pi)^2}q^{2m-2}e^{-i\frac{q^2}{2w}+i\bm{q}\cdot\bm{y}}\,.
\end{eqnarray}
Since the integral is dominated by the region $q\lesssim\sqrt{w}\,$, the factor $e^{i\bm{q}\cdot\bm{y}}$ can therefore be expanded, yielding
\begin{eqnarray}
    f_{m}(w,y)=2iw\frac{(-1)^m}{(2^{m}m!)^2}\sum_{k=0}^\infty\frac{(-y^2)^k}{(2^kk!)^2}\FP_{\epsilon\to0}\int_{\epsilon<|\bm{q}|}\frac{d^2q}{(2\pi)^2}q^{2m+2k-2}e^{-i\frac{q^2}{2w}}\,,
\end{eqnarray}
where we have used Eq. (\ref{Eq: theta integration}). By using Eq. (\ref{Eq: fresnel type integral}), $f_{m}(w,y)$ can be further expanded as
\begin{eqnarray}
    f_{m}(w,y)=
    \begin{dcases}
        \frac{i}{2\pi}w[\ln(-2iw)-\gamma]+\sum_{k=1}^\infty\frac{1}{\pi kk!}\qty(\frac{iw}{2})^{k+1}y^{2k}&(m=0)\\
        \sum_{k=0}^\infty\frac{(m+k-1)!}{\pi(m!k!)^2}\qty(\frac{iw}{2})^{m+k+1}y^{2k}&(m\geq1)
    \end{dcases}\,.
\end{eqnarray}

Next, $g_{s}(w,y)$ is defined for $s>0$ by
\begin{eqnarray}
    g_s(w,y)=2iw\FP_{\epsilon\to0}\int_{\epsilon<|\bm{q}|}\frac{d^2q}{(2\pi)^2}\frac{\chi_s(q)}{q^2}e^{-i\frac{q^2}{2w}+i\bm{q}\cdot\bm{y}}\,,
\end{eqnarray}
where $\chi_s(q)$ is defined by Eq. (\ref{Eq: chi definition}). As in the case of $f_{m}(w,y)\,$, $g_s(w,y)$ can be expanded as
\begin{eqnarray}
    g_s(w,y)=2iw\sum_{k=0}^\infty\frac{(-y^2)^k}{(2^kk!)^2}\FP_{\epsilon\to0}\int_{\epsilon<|\bm{q}|}\frac{d^2q}{(2\pi)^2}q^{2k-2}\chi_s(q)e^{-i\frac{q^2}{2w}}\,.\label{Eq: g_s expansion intermediate}
\end{eqnarray}
When $s\notin2\mathbb{N}\,$, by using Eq. (\ref{Eq: fresnel type integral}), we obtain
\begin{eqnarray}
    \FP_{\epsilon\to0}\int_{\epsilon<|\bm{q}|}\frac{d^2q}{(2\pi)^2}q^{2k-2}\chi_s(q)e^{-i\frac{q^2}{2w}}=\frac{\Gamma(1-s/2)\Gamma(s/2+k-1)}{\Gamma(s/2)}\qty(-\frac{iw}{2})^{s/2}(-2iw)^{k-1}\,.\nonumber\\
\end{eqnarray}
On the other hand, when $s=2l$ for $l\in\mathbb{N}\,$, using Eqs. (\ref{Eq: fresnel type integral}) and (\ref{Eq: fresnel type integral log}) yields
\begin{eqnarray}
    &&\FP_{\epsilon\to0}\int_{\epsilon<|\bm{q}|}\frac{d^2q}{(2\pi)^2}q^{2k-2}\chi_{2l}(q)e^{-i\frac{q^2}{2w}}\nonumber\\
    &&=
    \begin{dcases}
        -\frac{1}{2}\qty{\qty[\frac{\ln\qty(-iw/2)+\gamma}{2}]^2+\frac{\pi^2}{24}-\qty(\ln2-\gamma)^2}&(k=0\,,\,l=1)\\
        \frac{(l+k-2)!}{\qty[(l-1)!]^2}\qty(\frac{iw}{2})^l(-2iw)^{k-1}\\
        \phantom{AAA}\times\qty[\ln\qty(-\frac{iw}{2})+\gamma+H_{l+k-2}-2H_{l-1}]&(\text{otherwise})
    \end{dcases}\,.
\end{eqnarray}
Substituting these equations into Eq. (\ref{Eq: g_s expansion intermediate}), we finally obtain the expansion of $g_s(w,y)$ as
\begin{eqnarray}
    g_{s}(w,y)=-\sum_{k=0}^\infty\frac{\Gamma(1-s/2)\Gamma(s/2+k-1)}{\Gamma(s/2)(k!)^2}\qty(-\frac{iw}{2})^{s/2}\qty(\frac{i}{2}wy^2)^{k}\quad(s\notin\mathbb{N})\,,\label{Eq: g_s expansion}
\end{eqnarray}
\begin{eqnarray}
    g_{2l}(w,y)=
    \begin{dcases}
        -iw\qty{\qty[\frac{\ln\qty(-iw/2)+\gamma}{2}]^2+\frac{\pi^2}{24}-\qty(\ln2-\gamma)^2}\\
        \phantom{ -\sum_{k=0}^\infty}-\sum_{k=1}^\infty\frac{1}{kk!}\qty(\frac{iw}{2})^{k+1}y^{2k}\qty[\ln\qty(-\frac{iw}{2})+\gamma+H_{k-1}]&(l=1)\\
        -\sum_{k=0}^\infty\frac{(l+k-2)!}{\qty[(l-1)!k!]^2}\qty(\frac{iw}{2})^{k+l}y^{2k}\\
        \phantom{ -\sum_{k=0}^\infty}\times\qty[\ln\qty(-\frac{iw}{2})+\gamma+H_{l+k-2}-2H_{l-1}]&(l\geq2)
    \end{dcases}\,.
\end{eqnarray}
%

%===============================================

\bibliography{ref}

\end{document}